\documentclass{article}
\usepackage{amsfonts}
\usepackage{makeidx}
\usepackage{latexsym,amsmath,amssymb,amscd}
\usepackage{makecell}
\usepackage{xcolor}
\usepackage{multirow}
\usepackage[all]{xy}

\allowdisplaybreaks
\newtheorem{theorem}{Theorem}

\newtheorem{proposition}[theorem]{Proposition}

\begin{document}

\title{Quadratic first integrals of constrained autonomous conservative dynamical systems with fixed energy}
\author{Antonios Mitsopoulos$^{1,a)}$ and Michael Tsamparlis$^{2,3,b)}$ \\
{\ \ }\\
$^{1}${\textit{Faculty of Physics, Department of
Astronomy-Astrophysics-Mechanics,}}\\
{\ \textit{University of Athens, Panepistemiopolis, Athens 157 83, Greece}}
\\
$^{2}${\textit{NITheCS, National Institute for Theoretical and Computational
Sciences,}}\\
{\textit{KwaZulu-Natal, South Africa}}\\
$^{3}${\textit{TCCMMP, Theoretical and Computational Condensed Matter and
Materials Physics Group, }}\\
{\textit{School of Chemistry and Physics, }}\\
{\textit{University of KwaZulu-Natal, Pietermaritzburg, South Africa}}\\
\vspace{12pt} 
\\
$^{a)}$Author to whom correspondence should be addressed:
antmits@phys.uoa.gr 
\\
$^{b)}$Email: mtsampa@phys.uoa.gr \\
}
\date{}
\maketitle

\begin{abstract}
We consider autonomous conservative dynamical systems which are constrained with the condition that the total energy of the system has a specified value. We prove a theorem which provides the quadratic first integrals (QFIs), time-dependent and autonomous, of these systems in terms of the symmetries (conformal Killing vectors and conformal Killing tensors) of the kinetic metric. It is proved that there are three types of QFIs and for each type we give explicit formulae for their computation. It is also shown that when the autonomous QFIs are considered, then we recover the known results of previous works. For zero potential function, we have the case of constrained geodesics and obtain formulae to compute their QFIs. The theorem is applied in two cases. In the first case, we determine potentials which admit the second of the three types of QFIs. We recover a superintegrable potential of
the Ermakov type and a new integrable potential whose trajectories for zero energy and zero QFI are circles. In the second case, we integrate the constrained geodesic equations for a family of two-dimensional conformally flat metrics.
\end{abstract}


\section{Introduction}

We consider autonomous conservative dynamical systems of the form
\begin{equation}
\ddot{q}^{a}= -\Gamma^{a}_{bc}(q) \dot{q}^{b}\dot{q}^{c} -V^{,a}(q) \label{eq.cons1}
\end{equation}
which are subjected to the fixed energy constraint
\begin{equation}
\frac{1}{2}\gamma_{ab}(q)\dot{q}^{a}\dot{q}^{b} +V(q) = E_{0} \label{eq.cons2}
\end{equation}
where $q^{a}$ with $a=1,2,...,n$ are the generalized coordinates of the configuration space of the system, $n$ is the dimension of the configuration space, a dot over a letter indicates derivation with respect to (wrt) the parameter $t$ (time) along the trajectory $q^{a}(t)$, a comma denotes partial derivative, $\Gamma^{a}_{bc}(q)$ are the Riemannian connection coefficients defined by the kinetic metric $\gamma_{ab}(q)$ of the system, $V(q)$ is the potential function of the system and $E_{0}$ is a fixed constant (i.e. the total energy --Hamiltonian-- of the system). Moreover, Einstein's summation convention is applied and the kinetic metric $\gamma_{ab}(q)$ is used for lowering/raising the tensorial indices.

This type of constrained systems is of particular interest in many areas of Physics. For example, in Astrophysics and Celestial Mechanics, autonomous conservative systems with a fixed energy are used in order to model the galactic motion. In such studies (see e.g. \cite{Contopoulos 1960, Lynden 1962, Contopoulos 1963, Henon 1964, Contopoulos 2020}), it is assumed that the galaxy is described by an axisymmetric autonomous potential and the authors look for a third first integral (FI) in addition to the well-known FIs of energy and angular momentum. The problem is still open; however, it seems that additional FIs may exist in certain regions of the energy domain. Similar considerations have been done in non-relativistic Quantum Mechanics. Furthermore, in General Relativity, equations (\ref{eq.cons1}) and (\ref{eq.cons2}) with $V=0$ define constrained geodesic trajectories (timelike, spacelike, or null; depending upon the value of $E_{0}$) in a Riemannian spacetime which have numerous applications \cite{Eisenhart 1929, Benn 2006}.

The most important role of the FIs is the assessment of the integrability of a dynamical system, constrained or not. Therefore, it is important that systematic methods are developed which will allow the determination of FIs. Concerning the case of the non-quantum constrained dynamical systems, there
have been developed two types of such methods: a. The method which uses the dynamical equations directly, and b. The mini-superspace Lagrangian description. Each method follows a different approach, and its suitability depends on the particular application considered. The general description of each of these approaches has as follows.

\subsection*{a. The dynamical equations method}

There are three major approaches in this method which are the following.

\subsubsection*{The geodesic approach}

Because the $m$th-order polynomial in velocities FIs of the geodesic equations in a Riemannian space are known (see e.g. \cite{Katzin 1981, Mits 2021}), one transfers the dynamical equations to the geodesic equations of another Riemannian space. One approach is to introduce two extra degrees of freedom in the
configuration space. This is done by treating the time $t$ as a generalized coordinate $q^{n+1}=t$ and by introducing an additional dimension $q^{n+2}$ via the fixed energy constraint. This approach has been originated by Eisenhart \cite{Eisenhart 1929} who defined, in the $(n+2)$-dimensional expanded configuration space, the so-called Eisenhart metric whose geodesic equations produce both the dynamical equations and the energy constraint. A different approach is the Jacobi geometrization procedure (see e.g. \cite{Pin 1975, Abraham book}). According to this method, instead of introducing new generalized coordinates, one uses an energy-dependent reparameterization of the parameter of the dynamical equations to define a new metric, the Jacobi metric, whose timelike geodesics produce the dynamical equations and the constraint. The new metric is conformally related to the kinetic metric of the system with conformal factor the scalar quantity $E_{0} -V(q)$.

The approach of the Jacobi metric has been applied in \cite{Rosquist 1995, Karlovini 2000, Karlovini 2002, Pucacco 2005} in order to assess the integrability of two-dimensional (2d) autonomous conservative systems constrained at fixed and arbitrary energy. In particular, in \cite{Karlovini 2000, Karlovini 2002}, the integrable and superintegrable 2d Newtonian potentials that admit additional autonomous cubic and quartic FIs have been determined. However, certain drawbacks exist concerning the Jacobi metric when one increases the number of degrees of freedom, or the order of the FIs. This is due to the fact that: i) The conformal Killing vectors (CKVs) and the conformal Killing tensors (CKTs) of a general metric are not easy to compute, and ii) The Jacobi metric assumes a reparameterization from the `physical time' of the system to the so-called `Jacobi time' which affects the physical interpretation of the results.

\subsubsection*{The Lie symmetry approach}

This is a systematic and interesting approach which has been developed in \cite{Katzin 1968, Katzin 1973, Levine 1973} and has as follows.

One considers the unconstrained dynamical equations (\ref{eq.cons1}) and looks for Lie point symmetries generated by the infinitesimal point transformations:
\begin{equation}
\bar{q}^{a}(\bar{t})= q^{a}(t)+ \eta^{a}(q)\varepsilon, \enskip \bar{t}= t +\xi(t) \varepsilon \label{2.1}
\end{equation}%
where  $0 < \varepsilon \ll 1$, the generator $\xi(t)= 2\int \phi(q(t))dt \implies \dot{\xi}= 2\phi(q)$ and the function $\phi$ is evaluated along trajectories. These Lie point symmetries are referred to as trajectory collineations (TCs). It is proved that the generators $\eta^{a}(q)$ of the TCs are projective collineations (PCs) of the kinetic metric $\gamma_{ab}$ with projective function $\phi(q)$ which is related to the potential $V$ with the condition:
\begin{equation}
4\phi V^{,a} +\gamma^{ab}V_{;bc}\eta^{c} -\gamma^{bc} V_{,b} \eta^{a}{}_{;c}=0 \iff \mathcal{L}_{\boldsymbol{\eta}}V^{,a} +4\phi V^{,a}=0 \label{2.2}
\end{equation}%
where $\mathcal{L}_{\boldsymbol{\eta}}$ denotes the Lie derivative wrt $\eta^{a}$ (see also \cite{Tsamparlis 2011}). Furthermore, it is shown that the Lie derivative of a known FI (e.g. the total energy) wrt a TC is also a FI. This last result allows for a systematic evaluation of FIs once the TCs have been determined. These FIs have been called related FIs (RFIs). It is also shown that the quadratic RFIs share the same structure constants with the Lie algebra of TCs.

In \cite{Levine 1973}, the above approach was extended to the constrained autonomous conservative dynamical systems in the sense that one requires that the point transformations (\ref{2.1}) are constrained by the further requirement that the energy has a fixed value $E_{0}$. These new transformations map constrained solutions into constrained solutions and, hence, they are Lie point symmetries of the constrained system. These symmetries have been called natural trajectory collineations (NTCs). It is proved that the generators $\eta^{a}(q)$ of the NTCs are CKVs of the kinetic metric $\gamma_{ab}$ with conformal factor $\phi+c$, where $c$ is a constant, which are constrained by the new `selection rule'
\begin{equation}
V_{,a}\eta^{a} +2(\phi-c)(V-E_{0})=0. \label{2.3}
\end{equation}
It has been also shown that the formulation of the RFIs holds the same for the NTCs.

\subsubsection*{The direct approach}

In this approach, one assumes quadratic FIs (QFIs) of the functional form
\begin{equation}
I= K_{ab}(t,q)\dot{q}^{a}\dot{q}^{b} +K_{a}(t,q)\dot{q}^{a} +K(t,q)  \label{eq.cons3}
\end{equation}
where the coefficients are symmetric tensors depending on $q^{a}$ and $t$. The linear FIs (LFIs) are also included for $K_{ab}=0$.

The condition that (\ref{eq.cons3}) is a FI of the constrained system (\ref{eq.cons1}) - (\ref{eq.cons2}) is\footnote{Condition (\ref{eq.cons4}) is due to Hilbert's zero-theorem (see ch. IV, par. 8, pp. 166-167 in \cite{Hodge book}). According to this theorem, since $\frac{dI}{dt}$ is a third order polynomial in the velocities (after replacing $\ddot{q}^{a}$ from (\ref{eq.cons1}) ) and the constraint (\ref{eq.cons2}) is quadratic, there exists a first order ($3-2=1$) polynomial $\psi(t,q) +X_{c}(t,q)\dot{q}^{c}$ such that the quantity $\frac{dI}{dt}$ is expressed as in the condition (\ref{eq.cons4}).}
\begin{equation}
\frac{dI}{dt}= \left[ \psi(t,q) +X_{c}(t,q)\dot{q}^{c} \right] \left[ \gamma_{ab}\dot{q}^{a} \dot{q}^{b} +2(V-E_{0}) \right] \label{eq.cons4}
\end{equation}
where $\psi(t,q)$ is an arbitrary scalar and $X_{a}(t,q)$ is an arbitrary vector. The term $\psi(t,q) +X_{a}(t,q)\dot{q}^{a}$ plays the role of a `Lagrangian multiplier'. Using the dynamical equations (\ref{eq.cons1}) to replace the terms $\ddot{q}^{a}$ whenever they appear, condition (\ref{eq.cons4}) leads to a polynomial equation in the velocities $\dot{q}^{a}$.

At this point, there are two approaches. Either one looks for QFIs which hold for all values of the velocities on the constraint surface (we call them exact QFIs), or one looks for QFIs which are non-local, that is, they are valid only on trajectories specified by certain velocities. In the first approach, one sets all the coefficients of the powers of the velocities equal to zero and obtains a system of partial differential equations (PDEs) of the unknown quantities $K_{ab}, K_{a}, K$ and the dynamical quantities $V(q), E_{0}$. The solution of this system of PDEs provides the exact QFIs $I$. In the second approach, one does the same for all powers of the velocities which are of degree equal or higher than two, that is, the terms of zeroth and first order are excluded. In this case, one finds the conditional QFIs which are non-local QFIs, that is, they contain quantities within integrals. This is the type of FIs given, e.g., in \cite{Dimakis 2016, Dimakis 2019, Dimakis 2022}.

\subsection*{b. The mini-superspace Lagrangian method}

In this method, one introduces an extra dimension in the configuration space, the lapse function, and defines the mini-superspace Lagrangian of the system. The Euler-Lagrange (E-L) equations of this Lagrangian produce the dynamical equations and the equation of the constraint for $E_{0}=0$. This method is a particular case of a more general procedure applied on degenerate Lagrangian constrained systems. These are constrained systems whose Hessian vanishes, i.e. the quantity $\det \frac{\partial^{2}L}{\partial \dot{q}^{\alpha} \partial \dot{q}^{\beta}} =0$ where $q^{\alpha}$ denotes the generalized coordinates. In this case, a Hamiltonian cannot be defined because the Legendre transformation has no inverse. Since the
Hamiltonian is vital to Quantum Mechanics, the Dirac-Bergmann prescription has been developed \cite{Dirac 1950, Anderson 1951, Dirac 1958, Dirac 1964} to deal with this type of systems. In general terms, this formalism has as follows. The constraints are equations of the form $\phi_{A}(q,p)=0$ with $A=1,2,...,r$, where $r$ measures the degeneracy of the singular Lagrangian
(i.e. the degeneracy of the Hessian) and $p_{\alpha}= \frac{\partial L}{\partial \dot{q}^{\alpha}}$ are the conjugate momenta. Relations $\phi_{A}(q,p)=0$ are used to eliminate the $r$ degenerate coordinates. These relations are called
primary constraints and their time derivatives, secondary constraints.

Using the primary constraints, one splits the coordinates in two groups. The first group contains $n-r$ coordinates which define a non-vanishing Hessian, and the second group contains the remaining $r$ coordinates. This approach leads to conditional FIs which are non-local, that is, they are expressed in
terms of integrals and are generated by the CKVs and the CKTs instead of the Killing vectors (KVs) and the Killing tensors (KTs) which generate the exact QFIs. The conditional FIs can be used in the same way as the exact FIs; for example, in order to integrate the geodesic equations. A detailed and clear
discussion of this approach with applications can be found in \cite{Dimakis 2019}.
\bigskip

In the present work, we determine the QFIs (\ref{eq.cons3}) for the constrained dynamical systems (\ref{eq.cons1}) - (\ref{eq.cons2}) by using the direct method and the geometry generated by the kinetic metric of the system.

The structure of the paper is as follows.

In section \ref{sec.constrained.system}, we derive the main result of the present paper which is Theorem \ref{theorem.energy.constrained.systems}. It is shown that there are at most three types of exact QFIs associated with an autonomous conservative dynamical system constrained on a fixed energy level. Explicit formulae are given which provide these FIs in terms of the geometric symmetries of the kinetic metric. We note that these QFIs can be autonomous or time-dependent. In section \ref{sec.autonomous.FIs}, we consider the autonomous LFIs and QFIs derived from Theorem \ref{theorem.energy.constrained.systems}, and we recover the FIs found in \cite{Levine 1973}. In section \ref{sec.col.ckts2}, we recall the basic facts concerning the CKTs of order two, which are necessary in the computation of the QFIs of Theorem \ref{theorem.energy.constrained.systems}. In section \ref{sec.example1}, we apply Theorem \ref{theorem.energy.constrained.systems} to find 2d Newtonian potentials which admit a QFI of the second type $I_{(\ell)2}$. We consider the cases where the vector $L_{a}(q)$ is either a HV or a special CKV (SCKV). In the first case, we obtain a superintegrable potential of the Ermakov type and compute the trajectory for specific values of the constants. In the second case, we find a new class of integrable potentials and show that the orbits for $E_{0}=0$ and $I_{(0)2}=0$ are circles. In section \ref{sec.theorem.geodesics}, we specialize Theorem \ref{theorem.energy.constrained.systems} to the case of constrained geodesics and we collect our results in Theorem \ref{theorem.geodesics.constrained}. Applying Theorem \ref{theorem.geodesics.constrained}, we recover all previous results in the literature (see e.g. \cite{Katzin 1981, Tsamp 2020A, Tsamp 2020B}) for an arbitrary energy level $E_{0}$, and in Propositions \ref{pro.E.zero} and \ref{pro.E.nonzero} we determine the QFIs of null ($E_{0}=0)$ and non-null ($E_{0}\neq 0)$ geodesics, respectively. In section \ref{sec.example2}, we consider the non-null geodesics of a conformally flat metric and recover the results of \cite{Dimakis 2019} which were found using the rather complicated Dirac-Bergmann prescription. Finally, in section \ref{Conclusions}, we draw our conclusions.

\section{QFIs of constrained autonomous conservative dynamical systems with fixed energy}

\label{sec.constrained.system}

We consider autonomous conservative dynamical systems defined by equations (\ref{eq.cons1}) and (\ref{eq.cons2}), and we look for QFIs of the general form (\ref{eq.cons3}) defined by the condition (\ref{eq.cons4}).

Using the dynamical equations (\ref{eq.cons1}) to replace the quantities $\ddot{q}^{a}$ whenever they appear, condition (\ref{eq.cons4}) gives
\begin{align}
0= & \left( K_{(ab;c)} -X_{(a}\gamma_{bc)} \right) \dot{q}^{a}\dot{q}^{b}\dot{q}^{c} +\left( K_{ab,t} +K_{(a;b)} -\psi\gamma_{ab} \right) \dot{q}^{a}\dot{q}^{b} +\left[ K_{a,t}+ K_{,a} -2K_{ab}V^{,b} -2(V-E_{0})X_{a} \right] \dot{q}^{a} + \notag \\
& +K_{,t} -K_{a}V^{,a} -2(V-E_{0})\psi. \label{eq.cons4.1}
\end{align}
Assuming that there are no constraints on the velocities $\dot{q}^{a}$, we demand that the polynomial equation (\ref{eq.cons4.1}) is satisfied for all values of $\dot{q}^{a}$; therefore, the coefficient of each power of $\dot{q}^{a}$ must vanish. This leads to the system of PDEs:
\begin{eqnarray}
K_{(ab;c)} &=& X_{(a}\gamma_{bc)} \label{eq.cons5.1} \\
K_{(a;b)} &=& \psi\gamma_{ab} -K_{ab,t} \label{eq.cons5.2} \\
K_{,a} &=& 2K_{ab}V^{,b} +2(V-E_{0})X_{a} -K_{a,t} \label{eq.cons5.3} \\
K_{,t} &=& K_{a}V^{,a} +2(V-E_{0})\psi \label{eq.cons5.4}
\end{eqnarray}
where a semicolon denotes Riemannian covariant derivative.

Equation (\ref{eq.cons5.1}) implies that $K_{ab}$ is a CKT of order two of the kinetic metric $\gamma_{ab}$ with associated vector $X_{a}$. Contracting (\ref{eq.cons5.1}) with $\gamma^{bc}$, we find
\begin{equation}
X_{a}= \frac{1}{n+2} \left( K^{b}{}_{b;a} +2K^{b}{}_{a;b} \right). \label{eq.cons6.1}
\end{equation}

Contracting equation (\ref{eq.cons5.2}) with $\gamma^{ab}$, we obtain
\begin{equation}
\psi= \frac{1}{n} \left( K^{a}{}_{;a} +K^{a}{}_{a,t} \right). \label{eq.cons6.2}
\end{equation}

The system of PDEs (\ref{eq.cons5.1}) - (\ref{eq.cons5.4}) must be supplemented with the integrability conditions $K_{,[at]}=0$ and $K_{;[ab]}=0$ for the scalar $K$. We have:
\begin{eqnarray}
K_{a,tt} -2K_{ab,t}V^{,b} +\left( K_{b}V^{,b} \right)_{,a} +2(V-E_{0}) \left( \psi_{,a} -X_{a,t} \right) &=& 0 \label{eq.cons7.1} \\
2 \left( K_{[a|c|}V^{,c} \right)_{;b]} -K_{[a;b],t} +2\left[ (V-E_{0})X_{[a} \right]_{;b]} &=& 0 \label{eq.cons7.2}
\end{eqnarray}
where round (square) brackets indicate symmetrization (antisymmetrization) of the enclosed indices, and indices enclosed between vertical lines are overlooked by  (anti-)symmetrization symbols.

Finally, the system of PDEs which we have to solve consists of equations (\ref{eq.cons5.1}) - (\ref{eq.cons5.4}) and (\ref{eq.cons7.1}) - (\ref{eq.cons7.2}), where the quantities $X_{a}$ and $\psi$ are given by (\ref{eq.cons6.1}) and (\ref{eq.cons6.2}), respectively.

We state the solution of the above system of PDEs in Theorem \ref{theorem.energy.constrained.systems}. The proof is given in the Appendix.

\begin{theorem}
\label{theorem.energy.constrained.systems} The independent QFIs of the autonomous conservative dynamical system (\ref{eq.cons1}) subject to the fixed energy constraint (\ref{eq.cons2}) are the following:
\bigskip

\textbf{Integral 1.}
\begin{equation}
I_{(\ell)1}= \left( - \sum_{k=1}^{\ell} \frac{t^{2k}}{2k} L_{(2k-1)(a;b)} + C_{(0)ab}\right) \dot{q}^{a}\dot{q}^{b} + \sum_{k=1}^{\ell} t^{2k-1} L_{(2k-1)a}\dot{q}^{a} + \sum_{k=1}^{\ell} \frac{t^{2k}}{2k} L_{(2k-1)a}V^{,a}+ G(q)
\label{eq.cons19.1}
\end{equation}
where $C_{(0)ab}(q)$ and $L_{(2k-1)(a;b)}(q)$ for $k=1,2,...,\ell$ are CKTs with associated vectors $X_{(0)a}(q)$ and $Y_{(2k-1)a}(q)$, respectively, while the vectors $L_{(2k-1)a}(q)$ and the function $G(q)$ satisfy the conditions:
\begin{eqnarray}
\left( L_{(2\ell-1)b}V^{,b} \right)_{,a} &=& -2L_{(2\ell-1)(a;b)}V^{,b} -2(V-E_{0}) Y_{(2\ell-1)a} \label{eq.cons19.2} \\
\left( L_{(2k-1)b}V^{,b} \right)_{,a} &=& -2L_{(2k-1)(a;b)}V^{,b} -2k(2k+1)L_{(2k+1)a} -2(V-E_{0})Y_{(2k-1)a}, \enskip k=1,2, ..., \ell-1 \label{eq.cons19.3} \\
G_{,a}&=& 2C_{(0)ab}V^{,b} +2(V-E_{0}) X_{(0)a} -L_{(1)a}(\ell>0). \label{eq.cons19.4}
\end{eqnarray}

\textbf{Integral 2.}
\begin{equation}
I_{(\ell)2}= \sum_{k=0}^{\ell} \left( -\frac{t^{2k+1}}{2k+1} L_{(2k)(a;b)}\dot{q}^{a} \dot{q}^{b} +t^{2k} L_{(2k)a}\dot{q}^{a}+ \frac{t^{2k+1}}{2k+1} L_{(2k)a}V^{,a} \right) \label{eq.cons20.1}
\end{equation}
where $L_{(2k)(a;b)}(q)$ for $k=0,1,...,\ell$ are CKTs with associated vectors $Y_{(2k)a}(q)$ and the vectors $L_{(2k)a}(q)$ satisfy the conditions:
\begin{eqnarray}
\left( L_{(2\ell)b}V^{,b} \right)_{,a} &=& -2L_{(2\ell)(a;b)}V^{,b} -2(V-E_{0}) Y_{(2\ell)a} \label{eq.cons20.2} \\
\left( L_{(2k)b}V^{,b} \right)_{,a} &=& -2L_{(2k)(a;b)}V^{,b} -2(k+1)(2k+1)L_{(2k+2)a} -2(V-E_{0})Y_{(2k)a}, \enskip k=0,1,...,\ell-1. \label{eq.cons20.3}
\end{eqnarray}

\textbf{Integral 3.}
\begin{equation}
I_{(e)}= e^{\lambda t} \left( -L_{(a;b)} \dot{q}^{a} \dot{q}^{b} +\lambda L_{a}\dot{q}^{a} +L_{a}V^{,a} \right) \label{eq.cons21.1}
\end{equation}
where $\lambda\neq0$ is an arbitrary constant and $L_{(a;b)}(q)$ is a reducible CKT with associated vector $Y_{a}(q)$ such that
\begin{equation}
\left( L_{b}V^{,b} \right)_{,a}= -2L_{(a;b)}V^{,b} -\lambda^{2}L_{a} -2(V-E_{0})Y_{a}. \label{eq.cons21.2}
\end{equation}
\end{theorem}

\emph{Notation: In $I_{(\ell)\alpha}$ with $\alpha=1,2$, the index $(\ell)$ indicates the degree of the time-dependence, whereas the index $\alpha$ is used in order to distinguish between the two different types of independent QFIs. Moreover, in equation (\ref{eq.cons19.4}), the quantity $L_{(1)a}(\ell>0)$ indicates that the vector $L_{(1)a}$ exists only when $m>0$.}

We note that in the case that the involved second order CKTs of Theorem \ref{theorem.energy.constrained.systems} are KTs, the associated vectors $X_{(0)a}$ and $Y_{(N)a}$ vanish and Theorem \ref{theorem.energy.constrained.systems} reduces to Theorem 3 of \cite{Tsamp 2020B} for $Q^{a}=-V^{,a}$. Then, the associated independent QFIs hold for an arbitrary energy level $E_{0}$.

\subsection{The autonomous LFIs/QFIs of Theorem \ref{theorem.energy.constrained.systems}}

\label{sec.autonomous.FIs}

Theorem \ref{theorem.energy.constrained.systems} contains only two autonomous FIs. These are the following: \newline
a. The QFI\footnote{This is derived from the FI $I_{(\ell)1}$ given in (\ref{eq.cons19.1}) for time-dependence $\ell=0$. To simplify the notation, we set $C_{(0)ab}=C_{ab}$ and $X_{(0)ab}= X_{ab}$.}
\begin{equation}
J_{1}= C_{ab}(q)\dot{q}^{a}\dot{q}^{b} +G(q) \label{eq.cons22.1}
\end{equation}
where $C_{ab}$ is a second order CKT with associated vector $X_{a}(q)$ such that $G_{,a}= 2C_{ab}V^{,b} +2(V-E_{0})X_{a}$. \newline
b. The LFI
\begin{equation}
J_{2}= L_{a}(q)\dot{q}^{a} \label{eq.cons22.2}
\end{equation}
where $L_{a}$ is a CKV with conformal factor $\psi(q)$ such that $L_{a}V^{,a}= -2(V-E_{0})\psi$.

The LFI (\ref{eq.cons22.2}) is derived from the FI $I_{(\ell)2}$ given in (\ref{eq.cons20.1}) for time-dependence $\ell=0$ if we assume that $L_{(0)a}\equiv L_{a}(q)$ is a CKV with conformal factor $\psi(q)$. Indeed, we have
\[
L_{(a;b)}= \psi\gamma_{ab} \implies Y_{(0)a}= \psi_{,a}
\]
and the condition
\[
\left( L_{b}V^{,b} \right)_{,a} = -2\psi\gamma_{ab}V^{,b} -2(V-E_{0}) \psi_{,a} \implies L_{a}V^{,a} +2(V-E_{0})\psi =c =const.
\]
Then, the associated QFI becomes
\[
I_{(0)2}= t\left(-\psi\gamma_{ab}\dot{q}^{a}\dot{q}^{b} +L_{a}V^{,a} \right) +L_{a}\dot{q}^{a}= t\left[ 2(V-E_{0})\psi +L_{a}V^{,a} \right] +L_{a}\dot{q}^{a} = L_{a}\dot{q}^{a} +ct
\]
which for $c=0$ gives the autonomous LFI (\ref{eq.cons22.2}).

The autonomous FIs (\ref{eq.cons22.1}) and (\ref{eq.cons22.2}) are those found in sec. 6 of \cite{Levine 1973}. In particular: a. The QFI (\ref{eq.cons22.1}) coincides with eq. (6.5) of \cite{Levine 1973} and the additional conditions are eqs. (6.6) and (6.7); b. The LFI (\ref{eq.cons22.2}) coincides with eq. (6.1) of \cite{Levine 1973} and the additional conditions are eqs. (6.3) and (6.4). Moreover, the QFI given in eq. (4.7) of \cite{Terzis 2016} is derived from the QFI (\ref{eq.cons22.1}) for $G=E_{0}=0$. We note that the authors in \cite{Terzis 2016} determine this QFI by using the contact symmetries of the constrained system in the mini-superspace Lagrangian formalism.

\section{Conformal Killing tensors (CKTs) of order two}

\label{sec.col.ckts2}

In this section, we recall some basic results concerning the CKTs which will be used in applications of Theorem \ref{theorem.energy.constrained.systems} to be considered in the next sections.

A second order CKT \cite{Rani 2003} in an $m$-dimensional Riemannian manifold with local coordinates $x^{a}$ and metric $g_{ab}(x)$ is a symmetric tensor $U_{ab}$ such that
\begin{equation}
U_{(ab;c)} = u_{(a}g_{bc)} \iff U_{\{ab;c\}}= u_{\{a}g_{bc\}} \label{eq.ConKT.1}
\end{equation}
where $u_{a}$ is the vector associated to the CKT $U_{ab}$ and curly brackets denote cyclic permutation of the enclosed indices.

By contracting (\ref{eq.ConKT.1}) with $g^{ab}$, we find the vector
\begin{equation}
u_{a}= \frac{1}{m+2} \left( U_{;a} + 2U^{b}{}_{a;b} \right) \label{eq.ConKT.1a}
\end{equation}
where $U\equiv U^{a}{}_{a}$ is the trace of $U_{ab}$ and $m$ is the dimension of the manifold. The following terminology and results apply to CKTs: \newline
a. If $u_{a}=0$, $U_{ab}$ is a second order KT or an improper CKT. \newline
b. If $u_{a}\neq0$, $U_{ab}$ is called a proper CKT. \newline
c. If $u_{a}$ is a KV, $U_{ab}$ is called a homothetic KT (HKT) \cite{Prince 1983}. \newline
d. If the trace $U=0$, then $u_{a}= \frac{2}{m+2}U^{b}{}_{a;b}$ and $U_{ab}$ is called a trace-free CKT. \newline
e. If $u_{a}$ is a gradient (i.e. $u_{a}=u_{,a}$ where $u=u(x)$ is a scalar), then $U_{ab}$ is called a CKT of gradient type. \newline
f. If $f$ is an arbitrary function, then $\bar{U}_{ab}= U_{ab}+ fg_{ab}$ is a CKT with associated vector $\bar{u}_{a}= u_{a} + f_{;a}$. \newline
g. If $U_{ab}$ is a CKT of gradient type, then $C_{ab}= U_{ab} - ug_{ab}$ is a second order KT.

We can construct new CKTs as follows: \newline
i) If $f$ is an arbitrary function, then $fg_{ab}$ is a gradient CKT with associated vector $f_{,a}$.\newline
ii) If $X_{a}$ and $Y_{a}$ are CKVs with conformal factors, respectively, $\psi_{\mathbf{X}}$ and $\psi_{\mathbf{Y}}$, then the symmetrized tensor product $X_{(a}Y_{b)}$ is a CKT with associated vector $\psi_{X}Y_{a} +\psi_{Y}X_{a}$.\newline
iii) If $T_{ab}$ and $\bar{T}_{ab}$ are CKTs with associated vectors, respectively, $V_{a}$ and $W_{a}$, then the linear combination $\lambda T_{ab} + \mu \bar{T}_{ab}$, where $\lambda$ and $\mu$ are arbitrary constants, is a CKT with associated vector $\lambda V_{a} + \mu W_{a}$.

From i) - iii), we get the following general result \cite{Rani 2003, Walker 1970, Weir 1977}.

If an $m$-dimensional manifold admits $M$ CKVs $X_{Ka}$ with conformal factors $\psi_{K}$ where $K=1,2,...,M$, then
\begin{equation}
U_{ab} = fg_{ab} + c^{KL} X_{K(a}X_{|L|b)} \label{eq.ConKT.2}
\end{equation}
is a CKT with associated vector
\begin{equation}
u_{a}= f_{,a} + c^{KL} \left( \psi_{K}X_{La} + \psi_{L}X_{Ka} \right) \label{eq.ConKT.3}
\end{equation}
where $f$ is an arbitrary function, $c^{KL}$ are arbitrary constants and the summation is over the inequality $1 \leq K \leq L \leq M$. In flat spaces, all second order CKTs are of the form (\ref{eq.ConKT.2}).

\section{Example 1: Constrained orbits of potentials $V(x,y)$ in $E^{2}$ that admit FIs of the form $I_{(0)2}$ at fixed energy $E_{0}=0$}

\label{sec.example1}

In this case, the generalized coordinates $q^{a}=(x,y)$, the kinetic metric $\gamma_{ab}= \delta_{ab} =diag(1,1)$, the fixed energy level $E_{0}=0$ and the constraint (\ref{eq.cons2}) becomes
\begin{equation}
\gamma_{ab}\dot{q}^{a}\dot{q}^{b} +2V =0 \implies \dot{x}^{2} +\dot{y}^{2} +2V =0. \label{eq.cons.pot0}
\end{equation}

For time-dependence $\ell=0$ and fixed energy\footnote{We note that we can always insert the value of the constant $E_{0}$ into the potential; therefore, the case $E_{0}=0$ has no effect on the generality of our discussion.} $E_{0}=0$, the FI (\ref{eq.cons20.1}) becomes\footnote{To simplify the notation, we set $L_{(0)a}(q) \equiv L_{a}(q)$.}
\begin{equation}
I_{(0)2}= -tL_{(a;b)}\dot{q}^{a}\dot{q}^{b} +L_{a}\dot{q}^{a} +tL_{a}V^{,a} \label{eq.cons.pot1}
\end{equation}
where $L_{(a;b)}$ is a second order CKT of $\delta_{ab}$ with associated vector $Y_{a}$ such that
\begin{equation}
\left(L_{b}V^{,b}\right)_{,a}= -2L_{(a;b)}V^{,b} -2VY_{a}. \label{eq.cons.pot2}
\end{equation}

We consider various cases concerning the vector $L_{a}$.

\subsection{$L_{a}=(x,y)$ is the homothetic vector (HV)}

If the vector $L_{a}=(x,y)$ is the HV of $E^{2}$, then $L_{(a;b)}= \delta_{ab}$ and $Y_{a}=0$.

Using the constraint (\ref{eq.cons.pot0}), the QFI (\ref{eq.cons.pot1}) reduces to the LFI
\begin{equation}
I_{(0)2} \equiv I_{2}= -t\underbrace{(\dot{x}^{2} +\dot{y}^{2})}_{=-2V} +x\dot{x} +y\dot{y} +t (xV_{,x} +yV_{,y})= x\dot{x} +y\dot{y} +ct = r\dot{r} +ct \label{eq.cons.pot3}
\end{equation}
and the condition (\ref{eq.cons.pot2}) becomes
\begin{equation}
xV_{,x} +yV_{,y} +2V= c \label{eq.cons.pot4}
\end{equation}
where we have set $I_{(0)2} \equiv I_{2}$, $r^{2}= x^{2} +y^{2}$ and $c$ is an arbitrary constant.

Solving the PDE (\ref{eq.cons.pot4}), we find the potential \begin{equation}
V(x,y)= \frac{F\left(\frac{y}{x}\right)}{r^{2}} +\frac{c}{2} \label{eq.cons.pot5}
\end{equation}
where $F$ is an arbitrary smooth function of its argument.

The potential (\ref{eq.cons.pot5}) admits also the Ermakov QFI (see e.g. \cite{Hietarinta 1987, Mits 2dpots})
\begin{equation}
I_{1}= (x\dot{y} -y\dot{x})^{2} +2F;
\label{eq.cons.pot6}
\end{equation}
therefore, it is superintegrable. Using the FIs (\ref{eq.cons.pot3}) and (\ref{eq.cons.pot6}), the constraint (\ref{eq.cons.pot0}) becomes:
\begin{equation*}
\dot{x}^{2} +\dot{y}^{2} +\frac{2F}{r^{2}} +c =0 \implies (x\dot{x} +y\dot{y})^{2} +(x\dot{y} -y\dot{x})^{2} +2F +cr^{2} =0 \implies
\end{equation*}
\begin{equation}
(I_{2} -ct)^{2} +I_{1} +cr^{2} =0 \label{eq.cons.pot7}
\end{equation}
where $I_{1}$ and $I_{2}$ are arbitrary constants.

Integrating the LFI (\ref{eq.cons.pot3}), we find
\begin{equation}
\left( r^{2} \right)^{\cdot}= 2(I_{2} -ct) \implies r^{2}= -ct^{2} +2I_{2}t +c_{1} \implies r(t)= \sqrt{-ct^{2} +2I_{2}t +c_{1}} \label{eq.cons.pot8}
\end{equation}
where $c_{1}$ is an arbitrary constant.

Replacing (\ref{eq.cons.pot8}) in (\ref{eq.cons.pot7}), we get the condition
\begin{equation}
I_{1}= -I_{2}^{2} -cc_{1}. \label{eq.cons.pot9}
\end{equation}

Using polar coordinates $x=r\cos\theta$ and $y=r\sin\theta$, and conditions (\ref{eq.cons.pot8}) - (\ref{eq.cons.pot9}), the Ermakov QFI (\ref{eq.cons.pot6}) is written
\begin{equation}
r^{2}\dot{\theta}= \sqrt{-I_{2}^{2} -cc_{1} -2F} \implies \dot{\theta}= \frac{\sqrt{-I_{2}^{2} -cc_{1} -2F}}{-ct^{2} +2I_{2}t +c_{1}}. \label{eq.cons.pot10}
\end{equation}
Since $F$ is a function of $\frac{y}{x}= \tan\theta$, i.e. $F(\tan\theta)$, equation (\ref{eq.cons.pot10}) cannot be integrated to give $\theta(t)$.

In order to integrate (\ref{eq.cons.pot10}), we assume $c=0$ and $F=k=const$. Then, the potential (\ref{eq.cons.pot5}) reduces to the well-known Newton-Cotes potential (see e.g. \cite{Ibragimov 1998, Mits time})
\begin{equation}
V= \frac{k}{r^{2}} \label{eq.cons.pot11.1}
\end{equation}
and equations (\ref{eq.cons.pot8}), (\ref{eq.cons.pot10}) give the solution\footnote{It has been checked that the solution (\ref{eq.cons.pot11.2}) satisfies the polar Euler-Lagrange equation $\ddot{r} = r\dot{\theta}^{2} +\frac{2k}{r^{3}}$.}:
\begin{equation}
r(t)= \sqrt{2I_{2}t +c_{1}}, \enskip \theta(t)= \sqrt{-\frac{1}{4} -\frac{k}{2I_{2}^{2}}} \ln \left( 2I_{2}t +c_{1} \right) +\theta_{0} \label{eq.cons.pot11.2}
\end{equation}
which implies the orbit (with energy $E_{0}=0$!)
\begin{equation}
r= Ae^{B\theta} \label{eq.cons.pot11.3}
\end{equation}
where $\theta_{0}$ is an integration constant, $B\equiv \frac{1}{\sqrt{-1 -\frac{2k}{I_{2}^{2}}}}$ and $A \equiv e^{-B\theta_{0}}$.

The orbit (\ref{eq.cons.pot11.3}) is a logarithmic spiral (or miraculous spiral)  which has the following properties: a) For $B>0$, the size of the spiral increases outward as $\theta$ increases. b) For $B<0$, the size of the spiral decreases inward as $\theta$ increases.

We note that for the potential (\ref{eq.cons.pot11.1}), equation (\ref{eq.cons.pot10}) produces the LFI of the angular momentum (as expected for a central potential).

\subsection{$L_{a}$ is a special CKV (SCKV)}

It is well-known that $E^{2}$ admits two SCKVs: \newline
i) $B_{(1)a}=
\left(
  \begin{array}{c}
    \frac{x^{2}-y^{2}}{2} \\
    xy \\
  \end{array}
\right)$ with conformal factor $\psi_{1}= x$. \newline
ii) $B_{(2)a}=
\left(
  \begin{array}{c}
    xy \\
    \frac{y^{2}-x^{2}}{2} \\
  \end{array}
\right)$ with conformal factor $\psi_{2}= y$.

We assume that the vector $L_{a}= \left(
  \begin{array}{c}
    \frac{x^{2}-y^{2}}{2} \\
    xy \\
  \end{array} \right)$ is the SCKV $B_{(1)a}$ of $E^{2}$ with conformal factor $\psi=x$. Then, $L_{(a;b)}= x\delta_{ab}$ and $Y_{a}= \psi_{,a}= \left( 1, 0\right)$.

Using the constraint (\ref{eq.cons.pot0}), we find that the QFI (\ref{eq.cons.pot1}) reduces to the LFI
\begin{equation}
I_{(0)2}= -tx\underbrace{\left(\dot{x}^{2} +\dot{y}^{2}\right)}_{=-2V} +\frac{x^{2}-y^{2}}{2}\dot{x} +xy\dot{y} +t\left( \frac{x^{2}-y^{2}}{2} V_{,x} +xyV_{,y} \right)= \frac{x^{2}-y^{2}}{2}\dot{x} +xy\dot{y} +ct \label{eq.cons.pot12.1}
\end{equation}
and the associated condition (\ref{eq.cons.pot2}) becomes
\begin{equation}
\frac{x^{2}-y^{2}}{2} V_{,x} +xyV_{,y} +2xV =c \label{eq.cons.pot12.2}
\end{equation}
where $c$ is an arbitrary constant.

- For $c=0$, the PDE (\ref{eq.cons.pot12.2}) gives the potential
\begin{equation}
V= \frac{M\left( \frac{y}{r^{2}} \right)}{r^{4}} \label{eq.cons.pot12.3}
\end{equation}
where $r^{2}=x^{2}+y^{2}$ and $M$ is an arbitrary smooth function of its argument, the LFI (\ref{eq.cons.pot12.1}) becomes
\begin{equation}
I_{(0)2}= \frac{x^{2}-y^{2}}{2}\dot{x} +xy\dot{y} \label{eq.cons.pot12.4}
\end{equation}
and the constraint (\ref{eq.cons.pot0}) is written
\begin{equation}
\dot{x}^{2} +\dot{y}^{2} +\frac{2M}{r^{4}} =0. \label{eq.cons.pot12.5}
\end{equation}

We can use the LFI (\ref{eq.cons.pot12.4}) and the zero energy constraint (\ref{eq.cons.pot12.5}) in order to find an orbit with $E_{0}=0$ for the integrable potential (\ref{eq.cons.pot12.3}).

First, we assume the additional requirement $I_{(0)2}=0$. Then, the LFI (\ref{eq.cons.pot12.4}) becomes
\begin{equation}
\frac{x^{2}-y^{2}}{2}\dot{x} +xy\dot{y} =0. \label{eq.cons.pot12.6}
\end{equation}

Using polar coordinates, equation (\ref{eq.cons.pot12.6}) gives
\begin{equation}
\cos\theta \dot{r} +r\sin\theta \dot{\theta} =0 \implies \left( \ln r\right)^{\cdot} = -\tan\theta \dot{\theta} \implies \ln r = -\int \tan\theta d\theta \implies r= c_{1}\cos\theta \label{eq.cons.pot13}
\end{equation}
where $c_{1}\neq0$ is an arbitrary constant. The orbits (\ref{eq.cons.pot13}) are circles with centre $(\frac{c_{1}}{2}, 0)$ and radius $\frac{|c_{1}|}{2}$. Indeed, we have
\[
r=c_{1}\cos\theta \implies r^{2} =c_{1}r\cos\theta \implies x^{2}+y^{2} =c_{1}x \implies \left( x -\frac{c_{1}}{2} \right)^{2} +y^{2} =\frac{c_{1}^{2}}{4}.
\]

In order to compute the $\theta(t)$, we use the constraint (\ref{eq.cons.pot12.5}) which is written
\begin{equation}
\dot{r}^{2} +r^{2} \dot{\theta}^{2} + \frac{2 M\left( \frac{\sin\theta}{r} \right)}{r^{4}} =0. \label{eq.cons.pot14}
\end{equation}
Replacing the orbit (\ref{eq.cons.pot13}) in (\ref{eq.cons.pot14}), we get
\begin{equation}
\dot{\theta}^{2} = -\frac{2 M\left( \frac{\tan \theta}{c_{1}} \right)}{c_{1}^{6}\cos^{4}\theta} \implies t -t_{0} = c_{1}^{3}\int \frac{\cos^{2}\theta d\theta}{\sqrt{-2 M\left( \frac{\tan \theta}{c_{1}} \right)}} \label{eq.cons.pot15}
\end{equation}
where $t_{0}$ is an integration constant.

Therefore, at the fixed energy $E_{0}=0$, we have found the new class of integrable potentials (\ref{eq.cons.pot12.3}) and we have shown that for $I_{(0)2}=0$ the orbits of these potentials are circles.

\section{The LFIs/QFIs of the constrained geodesic equations}

\label{sec.theorem.geodesics}

Another important area in which we apply Theorem \ref{theorem.energy.constrained.systems} is the determination of the LFIs and QFIs of the constrained geodesic equations in an $n$-dimensional Riemannian manifold with metric $\gamma_{ab}(q).$ In this case, $V=0$ and Theorem \ref{theorem.energy.constrained.systems} takes the following form.

\begin{theorem}
\label{theorem.geodesics.constrained} The independent QFIs of the geodesic equations
\begin{equation}
\ddot{q}^{a} +\Gamma^{a}_{bc}(q)\dot{q}^{b}\dot{q}^{c} =0 \label{eq.congeo1}
\end{equation}
subject to the quadratic constraint
\begin{equation}
\gamma_{ab}(q)\dot{q}^{a}\dot{q}^{b}= 2E_{0} \label{eq.congeo2}
\end{equation}
where $\Gamma^{a}_{bc}$ are the Riemannian connection coefficients defined by the metric $\gamma_{ab}(q)$ and $E_{0}$ is an arbitrary fixed constant, are the following:
\bigskip

\textbf{Integral 1.}
\begin{equation}
I_{(\ell)1}= \left( - \sum_{k=1}^{\ell} \frac{t^{2k}}{2k} L_{(2k-1)(a;b)} + C_{(0)ab}\right) \dot{q}^{a}\dot{q}^{b} + \sum_{k=1}^{\ell} t^{2k-1} L_{(2k-1)a}\dot{q}^{a} + G(q)
\label{eq.congeo3.1}
\end{equation}
where $C_{(0)ab}(q)$ and $L_{(2k-1)(a;b)}(q)$ for $k=1,2,...,\ell$ are CKTs with associated vectors $X_{(0)a}(q)$ and $Y_{(2k-1)a}(q)$, respectively, while the vectors $L_{(2k-1)a}(q)$ and the function $G(q)$ satisfy the conditions:
\begin{eqnarray}
E_{0}Y_{(2\ell-1)a} &=& 0 \label{eq.congeo3.2} \\
E_{0}Y_{(2k-1)a} &=& k(2k+1)L_{(2k+1)a}, \enskip k=1,2, ..., \ell-1 \label{eq.congeo3.3} \\
G_{,a}&=& -2E_{0}X_{(0)a} -L_{(1)a}(\ell>0). \label{eq.congeo3.4}
\end{eqnarray}

\textbf{Integral 2.}
\begin{equation}
I_{(\ell)2}= \sum_{k=0}^{\ell} \left( -\frac{t^{2k+1}}{2k+1} L_{(2k)(a;b)}\dot{q}^{a} \dot{q}^{b} +t^{2k} L_{(2k)a}\dot{q}^{a} \right) \label{eq.congeo4.1}
\end{equation}
where $L_{(2k)(a;b)}(q)$ for $k=0,1,...,\ell$ are CKTs with associated vectors $Y_{(2k)a}(q)$ and the vectors $L_{(2k)a}(q)$ satisfy the conditions:
\begin{eqnarray}
E_{0}Y_{(2\ell)a} &=& 0 \label{eq.congeo4.2} \\
E_{0}Y_{(2k)a} &=& (k+1)(2k+1)L_{(2k+2)a}, \enskip k=0,1,...,\ell-1. \label{eq.congeo4.3}
\end{eqnarray}

\textbf{Integral 3.}
\begin{equation}
I_{(e)}= e^{\lambda t} \left( -L_{(a;b)} \dot{q}^{a} \dot{q}^{b} +\lambda L_{a}\dot{q}^{a} \right) \label{eq.congeo5.1}
\end{equation}
where $\lambda\neq0$ is an arbitrary constant and $L_{(a;b)}(q)$ is a reducible CKT with associated vector $Y_{a}(q)$ such that
\begin{equation}
L_{a}= \frac{2E_{0}}{\lambda^{2}}Y_{a}. \label{eq.congeo5.2}
\end{equation}
\end{theorem}

When there is no constraint of the form (\ref{eq.congeo2}), that is, the value of $E_{0}$ is arbitrary, the CKTs of Theorem \ref{theorem.geodesics.constrained} reduce to KTs (i.e. the associated vectors $X_{(0)a}= Y_{(k)a}=0$) and, as expected, the QFIs of Theorem \ref{theorem.geodesics.constrained} produce the well-known ones (see \cite{Katzin 1981}, Table IV of \cite{Tsamp 2020A}, and sec. X, par. A in \cite{Tsamp 2020B}).

\subsection{The QFIs for null constrained geodesic equations: Case $E_{0}=0$}

Applying Theorem \ref{theorem.geodesics.constrained} for $E_{0}=0$, we find the following result.

\begin{proposition}
\label{pro.E.zero} The independent QFIs of the null geodesic equations
\begin{equation}
\ddot{q}^{a} +\Gamma^{a}_{bc}(q)\dot{q}^{b}\dot{q}^{c} =0, \label{eq.congeo7}
\end{equation}
that is, subject to the constraint
\begin{equation}
\gamma_{ab}(q)\dot{q}^{a}\dot{q}^{b}= 0 \label{eq.congeo8}
\end{equation}
where $\Gamma^{a}_{bc}$ are the Riemannian connection coefficients defined by the metric $\gamma_{ab}(q)$, are the following:
\begin{equation}
I_{1}= C_{ab}\dot{q}^{a}\dot{q}^{b}, \enskip I_{2}= \frac{t^{2}}{2}G_{;ab} \dot{q}^{a}\dot{q}^{b} -tG_{,a}\dot{q}^{a} +G(q), \enskip I_{3}= -tL_{(a;b)} \dot{q}^{a}\dot{q}^{b} +L_{a}\dot{q}^{a} \label{eq.congeo9}
\end{equation}
where the quantities $C_{ab}$, $G_{;ab}$ and $L_{(a;b)}$ are second order CKTs of $\gamma_{ab}$.
\end{proposition}

Comparing the QFIs (\ref{eq.congeo9}) for null geodesics with the three QFIs of Table IV in \cite{Tsamp 2020A} for unconstrained geodesics, we may be tempted to say that they are the same. However, such a claim is not correct! Observe that in the former case the involved second order symmetric tensors are CKTs; whereas, in the latter case, they are KTs and the corresponding QFIs hold for an arbitrary $E_{0}$. This fundamental difference lies in the fact that the null geodesics are subject to the constraint (\ref{eq.congeo8}).

\subsection{The QFIs of non-null (spacelike or timelike) constrained geodesic equations: Case $E_{0}\neq0$}

Applying Theorem \ref{theorem.geodesics.constrained} for $E_{0}\neq0$, we have the following result.

\begin{proposition}
\label{pro.E.nonzero} The independent QFIs of the geodesic equations
\begin{equation}
\ddot{q}^{a} +\Gamma^{a}_{bc}(q)\dot{q}^{b}\dot{q}^{c} =0 \label{eq.congeo10.1}
\end{equation}
subject to the quadratic constraint
\begin{equation}
\gamma_{ab}(q)\dot{q}^{a}\dot{q}^{b}= 2E_{0} \neq0 \label{eq.congeo10.2}
\end{equation}
where $\Gamma^{a}_{bc}$ are the Riemannian connection coefficients defined by the metric $\gamma_{ab}(q)$ and $E_{0}\neq0$ is an arbitrary fixed constant, are the following:
\bigskip

\textbf{Integral 1.}
\begin{equation}
I_{(\ell)1}= \left( - \sum_{k=1}^{\ell} \frac{t^{2k}}{2k} L_{(2k-1)(a;b)} + C_{(0)ab}\right) \dot{q}^{a}\dot{q}^{b} + \sum_{k=1}^{\ell} t^{2k-1} L_{(2k-1)a}\dot{q}^{a} + G(q)
\label{eq.congeo11.1}
\end{equation}
where $L_{(2\ell-1)(a;b)}(q)$ is a KT, $C_{(0)ab}(q)$ and $L_{(2k-1)(a;b)}(q)$ for $k=1,2,...,\ell-1$ are CKTs with associated vectors $X_{(0)a}(q)$ and $Y_{(2k-1)a}= \frac{k(2k+1)}{E_{0}}L_{(2k+1)a}$, respectively, and the function $G(q)$ is such that
\begin{equation}
G_{,a}= -2E_{0}X_{(0)a} -L_{(1)a}(\ell>0). \label{eq.congeo11.2}
\end{equation}

\textbf{Integral 2.}
\begin{equation}
I_{(\ell)2}= \sum_{k=0}^{\ell} \left( -\frac{t^{2k+1}}{2k+1} L_{(2k)(a;b)}\dot{q}^{a} \dot{q}^{b} +t^{2k} L_{(2k)a}\dot{q}^{a} \right) \label{eq.congeo12}
\end{equation}
where $L_{(2\ell)(a;b)}$ is a KT and $L_{(2k)(a;b)}(q)$ for $k=0,1,...,\ell-1$ are CKTs with associated vectors $Y_{(2k)a}= \frac{(k+1)(2k+1)}{E_{0}}L_{(2k+2)a}$.

\textbf{Integral 3.}
\begin{equation}
I_{(e)}= e^{\lambda t} \left( -L_{(a;b)} \dot{q}^{a} \dot{q}^{b} +\lambda L_{a}\dot{q}^{a} \right) \label{eq.congeo13}
\end{equation}
where $\lambda\neq0$ is an arbitrary constant and $L_{(a;b)}(q)$ is a reducible CKT with associated vector $Y_{a}= \frac{\lambda^{2}}{2E_{0}}L_{a}$.
\end{proposition}

In the following section, we apply the results of section \ref{sec.theorem.geodesics} to a number of cases considered in \cite{Dimakis 2019}, and we show that we recover the answers without using the complex arguments of the mini-superspace Lagrangian method and the conditional FIs.

\section{Example 2: The non-null constrained geodesic equations of the metric $\gamma_{ab}= f(x,y)
\left(
  \begin{array}{cc}
    0 & 1 \\
    1 & 0 \\
  \end{array}
\right)$}

\label{sec.example2}

Consider the constrained dynamical system of two degrees of freedom $q^{a}=(x,y)$, which is described by the dynamical equations:
\begin{eqnarray}
\ddot{x} +\frac{f_{,x}}{f}\dot{x}^{2} &=& 0 \label{eq.2d2.1} \\
\ddot{y} +\frac{f_{,y}}{f}\dot{y}^{2} &=& 0 \label{eq.2d2.2} \\
f(x,y) \dot{x} \dot{y} &=& E_{0}. \label{eq.2d2.3}
\end{eqnarray}
Equations (\ref{eq.2d2.1}) and (\ref{eq.2d2.2}) are subjected to the quadratic constraint (\ref{eq.2d2.3}), where $f(x,y)$ is an arbitrary smooth function and $E_{0}$ is a non-zero fixed constant.

From the constraint equation (\ref{eq.2d2.3}), we read the 2d metric
\begin{equation}
\gamma_{ab}= f(x,y)
\left(
  \begin{array}{cc}
    0 & 1 \\
    1 & 0 \\
  \end{array}
\right). \label{eq.2d1}
\end{equation}
We observe that the Riemann tensor is
\begin{equation}
R_{1212}= f_{,xy} -\frac{f_{,x}f_{,y}}{f}= \frac{R}{2}\gamma_{1212} \label{eq.r1}
\end{equation}
where $\gamma_{abcd}\equiv \gamma_{ac}\gamma_{bd} -\gamma_{ad}\gamma_{bc}$ and $\gamma_{1212}= -f^{2}$. However, the metric (\ref{eq.2d1}) is not in general a metric of constant curvature because the Ricci scalar $R$ is not constant.

The non-vanishing Riemannian connection coefficients for the metric (\ref{eq.2d1}) are $\Gamma^{1}_{11}= \frac{f_{,x}}{f}$ and $\Gamma^{2}_{22}= \frac{f_{,y}}{f}$. We note then that the dynamical equations (\ref{eq.2d2.1}) and (\ref{eq.2d2.2}) are the geodesic equations (\ref{eq.congeo10.1}) of the metric (\ref{eq.2d1}) provided $t$ is the affine parameter. The constraint equation (\ref{eq.2d2.3}) is of the required form (\ref{eq.congeo10.2}).

The metric (\ref{eq.2d1}) has been discussed for various cases of the function $f(x,y)$ in \cite{Dimakis 2019}, using the rather complicated method of Dirac constraints \cite{Dirac 1958, Dirac 1964} and the associated conditional FIs. Our purpose here is to solve these cases, using the FIs of Proposition \ref{pro.E.nonzero}.

We shall need the CKVs of the metric (\ref{eq.2d1}) in order to compute from them the KVs and the CKTs of this metric. For the metric (\ref{eq.2d1}), we find the two-parameter family of CKVs
\begin{equation}
B_{a}= f(x,y)
\left(
  \begin{array}{c}
    F_{1}(y) \\
    F_{2}(x) \\
  \end{array}
\right) \implies \mathbf{B}= B^{a}\partial_{a}= F_{2}(x) \partial_{x} +F_{1}(y) \partial_{y} \label{eq.2d5}
\end{equation}
whose conformal factor
\begin{equation}
\psi(x,y)= \frac{F_{2} f_{,x} +F_{1} f_{,y} +f\left( \frac{dF_{1}}{dy} +\frac{dF_{2}}{dx} \right)}{2f}
\label{eq.2d5.1}
\end{equation}
where $F_{1}(y)$ and $F_{2}(x)$ are arbitrary smooth functions.

Using the general formulae (\ref{eq.ConKT.2}) and (\ref{eq.ConKT.3}), we construct for the metric (\ref{eq.2d1}) the second order CKT
\begin{equation}
C_{ab}= f^{2}(x,y)
\left(
  \begin{array}{cc}
    A_{1}(y) & 0 \\
    0 & A_{2}(x) \\
  \end{array}
\right) \label{eq.2d6}
\end{equation}
with associated vector
\begin{equation}
X_{a}=
\left(
  \begin{array}{c}
    f_{,y}A_{1}(y) +\frac{f}{2}\frac{dA_{1}}{dy} \\
    f_{,x}A_{2}(x) +\frac{f}{2}\frac{dA_{2}}{dx} \\
  \end{array}
\right) \label{eq.2d7}
\end{equation}
where $A_{1}(y)$ and $A_{2}(x)$ are arbitrary smooth functions.

To find the KVs, we require $\psi =0$. Then, condition (\ref{eq.2d5.1}) implies that
\begin{equation}
F_{2} f_{,x} +F_{1} f_{,y} +f\left( \frac{dF_{1}}{dy} +\frac{dF_{2}}{dx} \right) =0. \label{eq.2d7a}
\end{equation}

\subsection{The case of a space of constant curvature: LFIs}

We require the metric (\ref{eq.2d1}) to be a metric of constant curvature. Since the metric (\ref{eq.2d1}) is 2d and satisfies the condition (\ref{eq.r1}), it must also satisfy the additional requirement that $R=const$.

Without loss of generality, we set $R=-\frac{4}{k}$ where $k\neq0$ is an arbitrary non-zero constant. Replacing this $R$ in (\ref{eq.r1}), we get the non-linear PDE:
\begin{equation}
f_{,xy} -\frac{f_{,x}f_{,y}}{f} -\frac{2f^{2}}{k}= 0 \label{eq.r2}
\end{equation}
whose solution is $f(x,y)= \frac{k}{(x+y)^{2}}$. Therefore, the metric of constant curvature is
\begin{equation}
\gamma_{ab}= \frac{k}{(x+y)^{2}}
\left(
  \begin{array}{cc}
    0 & 1 \\
    1 & 0 \\
  \end{array}
\right). \label{eq.2d10}
\end{equation}

The constrained system (\ref{eq.2d2.1}) - (\ref{eq.2d2.3}) becomes:
\begin{eqnarray}
\ddot{x} -\frac{2}{x+y}\dot{x}^{2} &=& 0 \label{eq.2d11.1} \\
\ddot{y} -\frac{2}{x+y}\dot{y}^{2} &=& 0 \label{eq.2d11.2} \\
\frac{k}{(x+y)^{2}}\dot{x}\dot{y} &=& E_{0} \label{eq.2d11.3}
\end{eqnarray}

A space of constant curvature has three KVs. Therefore, the system of the constrained geodesic equations (\ref{eq.2d11.1}) - (\ref{eq.2d11.3}) admits three independent LFIs and, as a result, it is superintegrable. To find the KVs, we use equation (\ref{eq.2d7a}) which in this case becomes
\begin{equation}
x\frac{dF_{1}}{dy} +y\frac{dF_{2}}{dx} +\left( y\frac{dF_{1}}{dy} -2F_{1} \right) +\left( x\frac{dF_{2}}{dx} -2F_{2} \right) =0. \label{eq.2d21}
\end{equation}
From the leading terms $x\frac{dF_{1}}{dy}$ and $y\frac{dF_{2}}{dx}$ in (\ref{eq.2d21}), we deduce that the function $F_{1}(y)$ must be of the form
\begin{equation}
F_{1}(y)= c_{1}y^{2} +c_{2}y +c_{3} \label{eq.2d22}
\end{equation}
where $c_{1}, c_{2}, c_{3}$ are arbitrary constants.

Replacing (\ref{eq.2d22}) in (\ref{eq.2d21}), we find
\begin{equation}
F_{2}(x)= -c_{1}x^{2} +c_{2}x -c_{3}. \label{eq.2d22.1}
\end{equation}

Substituting the functions (\ref{eq.2d22}) and (\ref{eq.2d22.1}) in (\ref{eq.2d5}), we find for the metric (\ref{eq.2d10}) the following three KVs (one for each constant $c_{i}$, $i=1,2,3$):
\begin{equation}
L_{(1)a}= \frac{k}{(x+y)^{2}}
\left(
  \begin{array}{c}
    y^{2} \\
    -x^{2} \\
  \end{array}
\right), \enskip L_{(2)a}= \frac{k}{(x+y)^{2}}
\left(
  \begin{array}{c}
    y \\
    x \\
  \end{array}
\right), \enskip
L_{(3)a}= \frac{k}{(x+y)^{2}}
\left(
  \begin{array}{c}
    1 \\
    -1 \\
  \end{array}
\right). \label{eq.2d23}
\end{equation}

Replacing the KVs (\ref{eq.2d23}) in the FI (\ref{eq.congeo12}) for $\ell=0$, we find the following three LFIs:
\begin{equation}
I_{1}= \frac{k}{(x+y)^{2}} \left( y^{2}\dot{x} -x^{2}\dot{y} \right), \enskip I_{2}= \frac{k}{(x+y)^{2}} \left( y\dot{x} +x\dot{y} \right), \enskip I_{3}= \frac{k}{(x+y)^{2}} \left( \dot{x} -\dot{y} \right). \label{eq.2d24}
\end{equation}
These LFIs establish the superintegrability of the geodesic equations (\ref{eq.2d11.1}) - (\ref{eq.2d11.2}).

\subsubsection{Integration of the constrained geodesics (\ref{eq.2d11.1}) - (\ref{eq.2d11.3})}

To simplify the notation, we introduce the constants $a_{0}= \frac{E_{0}}{k}\neq0$ and $a_{D}= \frac{I_{D}}{k}$ with $D=1,2,3$.

Then, the fixed energy constraint (\ref{eq.2d11.3}) and the LFIs (\ref{eq.2d24}) are written:
\begin{eqnarray}
a_{0}(x+y)^{2}&=& \dot{x}\dot{y} \label{eq.2d25.1} \\
a_{1}(x+y)^{2}&=& y^{2}\dot{x} -x^{2}\dot{y} \label{eq.2d25.2} \\
a_{2}(x+y)^{2}&=& y\dot{x} +x\dot{y} \label{eq.2d25.3} \\
a_{3}(x+y)^{2}&=& \dot{x} -\dot{y}. \label{eq.2d25.4}
\end{eqnarray}

Solving equations (\ref{eq.2d25.3}) and (\ref{eq.2d25.4}) wrt $\dot{x}$ and $\dot{y}$, we find that:
\begin{eqnarray}
\dot{x} &=& (a_{2} +a_{3}x)(x+y) \label{eq.2d26.1} \\
\dot{y} &=& (a_{2} -a_{3}y)(x+y). \label{eq.2d26.2}
\end{eqnarray}

Replacing (\ref{eq.2d26.1}) and (\ref{eq.2d26.2}) in the remaining equations (\ref{eq.2d25.1}) and (\ref{eq.2d25.2}), we obtain, respectively, the following:
\begin{eqnarray}
a_{0}&=& (a_{2} -a_{3}y)(a_{2} +a_{3}x) \label{eq.2d26.3} \\
0&=& (a_{3}x +a_{2})y^{2} +(a_{3}x^{2} -a_{1})y -x(a_{2}x +a_{1}). \label{eq.2d26.4}
\end{eqnarray}

Solving the quadratic algebraic equation (\ref{eq.2d26.4}) wrt $y$, we find the solution\footnote{The solution $y=-x$ is not acceptable, because of the requirement $x+y \neq0$; otherwise, the function $f= \frac{k}{(x+y)^{2}}$ is not well-defined.}
\begin{equation}
y= \frac{a_{2}x +a_{1}}{a_{3}x +a_{2}} \label{eq.2d27}
\end{equation}
which when substituted into (\ref{eq.2d26.3}) gives the relation
\begin{equation}
a_{0}= a_{2}^{2} -a_{1}a_{3} \neq0 \label{eq.2d28}
\end{equation}
where $a_{0}$ is a non-zero constant fixed by the energy $E_{0}$ of the geodesics. This means that the constants $a_{1}, a_{2}$ and $a_{3}$ take only those values which satisfy the relation (\ref{eq.2d28}) as fixed by the `energy' constant $a_{0}$. We note also that the orbit (\ref{eq.2d27}) coincides with eq. (6.17) of \cite{Dimakis 2019}, if we rename the constants as follows: $a_{1}= \mp\kappa_{2}^{2} \sqrt{2\kappa_{1}^{2} -R}$, $a_{2}= \sqrt{2}\kappa_{1}\kappa_{2}$ and $a_{3}= \mp\sqrt{2\kappa_{1}^{2} -R}$.

Replacing the orbit (\ref{eq.2d27}) in (\ref{eq.2d26.1}), we find:
\begin{equation}
\dot{x}= a_{3}x^{2} +2a_{2}x +a_{1} \implies t= \int \frac{dx}{a_{3}x^{2} +2a_{2}x +a_{1}}. \label{eq.2d29}
\end{equation}
There are two cases to be considered.

a) Case $a_{3}=0$.

In this case, from equations (\ref{eq.2d27})and (\ref{eq.2d29}), we find the parametric geodesic solutions:
\begin{equation}
x(t)= c_{1}e^{2a_{2}t} +c_{2}, \enskip y(t)= c_{1}e^{2a_{2}t} -c_{2} \label{eq.2d30}
\end{equation}
where $c_{1}$ is an arbitrary constant and the constant $c_{2}\equiv -\frac{a_{1}}{2a_{2}}$. Equation (\ref{eq.2d28}) implies that the constant $a_{2}$ is fixed by the energy constraint $a_{0}= a_{2}^{2}\neq0$, while the remaining equation (\ref{eq.2d26.2}) is satisfied identically.

b) Case $a_{3}\neq0$.

In this case, from equations (\ref{eq.2d27})and (\ref{eq.2d29}), we find the parametric geodesic solutions:
\begin{equation}
x(t)= \frac{\sqrt{-a_{0}}}{a_{3}} \tan \left( \sqrt{-a_{0}}t +c_{0} \right) -\frac{a_{2}}{a_{3}}, \enskip y(t)=  \frac{\sqrt{-a_{0}}}{a_{3}} \frac{1}{\tan \left( \sqrt{-a_{0}}t +c_{0} \right)} +\frac{a_{2}}{a_{3}} \label{eq.2d31}
\end{equation}
where $c_{0}$ is an arbitrary constant and $a_{0}= a_{2}^{2} -a_{1}a_{3} <0$. From equation (\ref{eq.2d28}), the constant $a_{1}= \frac{a_{2}^{2}}{a_{3}} -\frac{a_{0}}{a_{3}}$, while the remaining equation (\ref{eq.2d26.2}) is satisfied identically.
\bigskip

\underline{Remark 1:} The choice $f(x,y)=x$ made in \cite{Dimakis 2019} (see sec. VI, eq. (6.1) ) leads to the flat Lorentzian metric and the solution of the geodesic equations is straightforward. Indeed, the coordinate transformation $q^{a}=(x,y) \to \bar{q}^{a}=(u,v)$ given by the relations:
\[
u= y -\frac{x^{2}}{4}, \enskip v=y +\frac{x^{2}}{4}
\]
brings the metric $\gamma_{ab}=
\left(
  \begin{array}{cc}
    0 & x \\
    x & 0 \\
  \end{array}
\right)$ to its canonical form $\eta_{ab}=diag(-1,+1)$.

In the coordinates $(u,v)$, the solution of the constrained geodesic equations is:
\[
u(t)= k_{1}t +k_{2}, \enskip v(t)= k_{3}t +k_{4}
\]
where $k_{1}, k_{2}, k_{3}$ and $k_{4}$ are arbitrary constants, while the associated quadratic constraint (\ref{eq.congeo10.2}) implies that
\[
\eta_{ab}\dot{\bar{q}}^{a}\dot{\bar{q}}^{b}= 2E_{0}\neq0 \implies -\dot{u}^{2} +\dot{v}^{2}= 2E_{0} \implies E_{0}= \frac{1}{2}\left( k_{3}^{2} -k_{1}^{2} \right).
\]
Therefore, in this case, the application of neither the exact nor the conditional FIs is necessary.

\subsection{The case of a metric that does not possess KVs: QFIs}

In this case, we shall use the QFI $I_{(\ell)1}$ given in (\ref{eq.congeo11.1}) for $\ell=0$, that is\footnote{To simplify the notation, we set $C_{(0)ab}=C_{ab}(q)$ and $X_{(0)a}=X_{a}(q)$.},
\begin{equation}
I_{(0)1}= C_{ab}(q)\dot{q}^{a}\dot{q}^{b} +G(q) \label{eq.2d3}
\end{equation}
where $C_{ab}$ is a CKT of the metric (\ref{eq.2d1}) with associated vector $X_{a}(q)$ and the function $G(q)$ is such that
\begin{equation}
G_{,a}= -2E_{0}X_{a}. \label{eq.2d4}
\end{equation}

Replacing $C_{ab}$ from (\ref{eq.2d6}) in (\ref{eq.2d3}), the QFI is written
\begin{equation}
I_{(0)1}= f^{2} \left[ A_{1}(y)\dot{x}^{2} +A_{2}(x)\dot{y}^{2} \right] +G(q). \label{eq.2d3.1}
\end{equation}

Replacing $X_{a}$ from (\ref{eq.2d7}) in the condition (\ref{eq.2d4}), we obtain the following system of PDEs:
\begin{equation}
G_{,x}= -2E_{0} \left[ f_{,y}A_{1}(y) +\frac{f}{2} \frac{dA_{1}}{dy} \right], \enskip G_{,y}= -2E_{0} \left[ f_{,x}A_{2}(x) +\frac{f}{2}\frac{dA_{2}}{dx} \right]. \label{eq.2d8}
\end{equation}

Taking the integrability condition $G_{,[xy]}=0$ of the function $G(x,y)$, we find the second order PDE:
\begin{equation}
f_{,yy}A_{1}(y) -f_{,xx}A_{2}(x) +\frac{3}{2}\left( f_{,y}\frac{dA_{1}}{dy} -f_{,x}\frac{dA_{2}}{dx}\right) +\frac{f}{2}\left( \frac{d^{2}A_{1}}{dy^{2}} -\frac{d^{2}A_{2}}{dx^{2}}\right) =0. \label{eq.2d9}
\end{equation}

Equation (\ref{eq.2d9}) cannot be solved in full generality because it has three unknown functions $f(x,y)$, $A_{1}(y)$ and $A_{2}(x)$; therefore, it is an overdetermined PDE. In order to find a solution, we must fix two of the three unknowns.

First, we choose $f(x,y)= -x^{3}e^{y}(x+e^{y})$.

For this choice, the metric (\ref{eq.2d1}) is (see eq. (6.26) of \cite{Dimakis 2019})
\begin{equation}
\gamma_{ab}= -x^{3}e^{y}(x+e^{y})
\left(
  \begin{array}{cc}
    0 & 1 \\
    1 & 0 \\
  \end{array}
\right) \label{eq.2d32}
\end{equation}
with Ricci scalar $R= -\frac{2}{x^{3}(x+e^{y})^{3}}$ and Riemann tensor $R_{1212}= \frac{x^{3}e^{2y}}{x+e^{y}}$. The metric (\ref{eq.2d32}) does not possess KVs (or non-trivial second order KTs) and, therefore, to assess the integrability of its corresponding constrained geodesics, we should focus to the proper CKVs and CKTs.

The constrained system (\ref{eq.2d2.1}) - (\ref{eq.2d2.3}) becomes:
\begin{eqnarray}
\ddot{x} +\left( \frac{1}{x+e^{y}} +\frac{3}{x} \right) \dot{x}^{2} &=& 0 \label{eq.2d33.1} \\
\ddot{y} +\left( \frac{e^{y}}{x+e^{y}} +1 \right)\dot{y}^{2} &=& 0 \label{eq.2d33.2} \\
-x^{3}e^{y}(x+e^{y})\dot{x}\dot{y} &=& E_{0}. \label{eq.2d33.3}
\end{eqnarray}

In order to solve (\ref{eq.2d9}), we have to fix either $A_{1}(y)$ or $A_{2}(x)$. We choose $A_{2}(x)=0$ and the PDE (\ref{eq.2d9}) gives the first degree polynomial in $x$
\begin{equation}
\left( A_{1} +\frac{3}{2}\frac{dA_{1}}{dy} +\frac{1}{2} \frac{d^{2}A_{1}}{dy^{2}} \right)x +e^{y} \left( 4A_{1} +3\frac{dA_{1}}{dy} +\frac{1}{2} \frac{d^{2}A_{1}}{dy^{2}} \right) =0 \label{eq.2d34}
\end{equation}
whose solution is\footnote{We note that if instead of $A_{2}(x)=0$ we had chosen $A_{1}(y)=0$, the PDE (\ref{eq.2d9}) would have given the trivial solution $A_{2}(x)=0$.} $A_{1}(y)= e^{-2y}$.

Having fixed the three unknown functions:
\begin{equation}
f(x,y)= -x^{3}e^{y}(x+e^{y}), \enskip A_{1}(y)= e^{-2y}, \enskip A_{2}(x)=0 \label{eq.2d35}
\end{equation}
the system of PDEs (\ref{eq.2d8}) becomes:
\begin{equation}
G_{,x}= 2E_{0}x^{3}, \enskip G_{,y}=0 \label{eq.2d36}
\end{equation}
from which we find the function
\begin{equation}
G= \frac{E_{0}}{2}x^{4}. \label{eq.2d37}
\end{equation}

Replacing (\ref{eq.2d35}) and (\ref{eq.2d37}) in (\ref{eq.2d3.1}), we find for the constrained non-null geodesic equations (\ref{eq.2d33.1}) - (\ref{eq.2d33.3}) the QFI
\begin{equation}
I_{1}=  x^{6}(x+e^{y})^{2}\dot{x}^{2} +\frac{E_{0}}{2}x^{4}. \label{eq.2d39}
\end{equation}

Using the QFI (\ref{eq.2d39}) and the quadratic constraint (\ref{eq.2d33.3}), the constrained non-null geodesics (\ref{eq.2d33.1}) - (\ref{eq.2d33.3}) can be integrated as follows.

First, we assume that $I_{1}=0$. Then, the QFI (\ref{eq.2d39}) gives
\begin{equation}
E_{0}= -2x^{2}(x+e^{y})^{2}\dot{x}^{2}. \label{eq.2d40}
\end{equation}
Replacing (\ref{eq.2d40}) in the constraint (\ref{eq.2d33.3}), we find the trajectory
\begin{equation}
\frac{\dot{y}}{\dot{x}}= \frac{2(x+e^{y})}{xe^{y}} \implies \frac{dy}{dx}= \frac{2(x+e^{y})}{xe^{y}} \implies y= \ln \left( c_{1}x^{2} -2x \right). \label{eq.2d41}
\end{equation}
where $c_{1}$ is a constant. This result coincides with eq. (6.27c) of \cite{Dimakis 2019}.

Moreover, replacing (\ref{eq.2d41}) in (\ref{eq.2d40}), we find
\begin{equation}
\left( c_{1}x^{3} -x^{2} \right)^{2} \dot{x}^{2}= -\frac{E_{0}}{2} >0 \label{eq.2d42}
\end{equation}
since $E_{0}\neq0$. Therefore, $E_{0}<0$ and equation (\ref{eq.2d42}) implies that
\begin{equation}
\int \left( c_{1}x^{3} -x^{2} \right) dx= \pm \sqrt{-\frac{E_{0}}{2}} \int dt \implies t= \pm \sqrt{-\frac{2}{E_{0}}} \left( \frac{c_{1}}{4}x^{4} -\frac{x^{3}}{3} \right) +t_{0} \label{eq.2d43}
\end{equation}
where $t_{0}$ is an integration constant.
\bigskip

\underline{Remark 2:} As we have seen, $t$ is the affine parameter of the geodesics. If we assume $E_{0}=-\frac{1}{2}$ and a reparameterization $t=t(\tau)$ such that in the new parameter $x(\tau)=\tau$, equation (\ref{eq.2d43}) determines the parameter transformation
\[
\frac{dt}{d\tau}= \pm 2\tau^{2} \left( c_{1}\tau -1 \right) \equiv N(\tau).
\]
The function $N(\tau)$ is the lapse function of the corresponding mini-superspace Lagrangian (see eq. (6.27a) of \cite{Dimakis 2019}).

\subsection{The case of a class of integrable Lorentzian Toda systems: QFIs}

Another case where QFIs are required is a class of 2d integrable Lorentzian Toda systems \cite{Gavrilov 1998} which is equivalent to the constrained dynamical system (\ref{eq.2d2.1}) - (\ref{eq.2d2.3}) for (see eq. (6.35) of \cite{Dimakis 2019})
\begin{equation}
f(x,y)= k_{1} e^{\sqrt{2} \left[ b_{1}x +(b_{2} -b_{1})y \right]} +k_{2} e^{\frac{1}{\sqrt{2}}\left( b_{1}x +b_{3}y \right)} \label{eq.2d44}
\end{equation}
where $k_{1}, k_{2}, b_{1}, b_{2}, b_{3}$ are arbitrary non-zero constants and $b_{1}\neq b_{2}$.

In this case, the metric (\ref{eq.2d1}) is
\begin{equation}
\gamma_{ab}= \left(k_{1} e^{\sqrt{2} \left[ b_{1}x +(b_{2} -b_{1})y \right]} +k_{2} e^{\frac{1}{\sqrt{2}}\left( b_{1}x +b_{3}y \right)} \right)
\left(
  \begin{array}{cc}
    0 & 1 \\
    1 & 0 \\
  \end{array}
\right). \label{eq.2d45}
\end{equation}
We note that this metric is in fact the Jacobi metric of the system, and it is flat only when $b_{3}= 2(b_{2}-b_{1})$.

In order to solve (\ref{eq.2d9}), we choose $A_{1}(y)=0$. Then, the PDE (\ref{eq.2d9}) gives $A_{2}(x)= e^{-\sqrt{2}b_{1}x}$.

Having fixed the three unknown functions:
\begin{equation}
f(x,y)= k_{1} e^{\sqrt{2} \left[ b_{1}x +(b_{2} -b_{1})y \right]} +k_{2} e^{\frac{1}{\sqrt{2}}\left( b_{1}x +b_{3}y \right)}, \enskip A_{1}(y)= 0, \enskip A_{2}(x)= e^{-\sqrt{2}b_{1}x} \label{eq.2d46}
\end{equation}
the system of PDEs (\ref{eq.2d8}) becomes:
\begin{equation}
G_{,x}= 0, \enskip G_{,y}=-\sqrt{2}k_{1}b_{1}E_{0} e^{\sqrt{2}(b_{2}-b_{1})y} \label{eq.2d47}
\end{equation}
from which we find the function
\begin{equation}
G= \frac{k_{1}b_{1}E_{0}}{b_{1}-b_{2}} e^{\sqrt{2}(b_{2}-b_{1})y}. \label{eq.2d48}
\end{equation}

Replacing (\ref{eq.2d46}) and (\ref{eq.2d48}) in (\ref{eq.2d3.1}), we find for the constrained non-null geodesic equations (\ref{eq.2d2.1}) - (\ref{eq.2d2.3}) with $f(x,y)$ given by (\ref{eq.2d44}) the QFI
\begin{equation}
I_{1}=  \left( k_{1} e^{\sqrt{2} \left[ b_{1}x +(b_{2} -b_{1})y \right]} +k_{2} e^{\frac{1}{\sqrt{2}}\left( b_{1}x +b_{3}y \right)} \right)^{2} e^{-\sqrt{2}b_{1}x} \dot{y}^{2} +\frac{k_{1}b_{1}E_{0}}{b_{1}-b_{2}} e^{\sqrt{2}(b_{2}-b_{1})y}. \label{eq.2d50}
\end{equation}

Using the QFI (\ref{eq.2d50}) and the corresponding quadratic constraint, the integration of the constrained non-null geodesics for $I_{1}=0$ is straightforward (see previous examples). We conclude that the conditional LFI used in \cite{Dimakis 2019} is again not necessary because we can solve the problem by using instead the exact QFI (\ref{eq.2d50}).

\section{Conclusions}

\label{Conclusions}

The assessment of the integrability of a dynamical system requires the knowledge of `enough' in number (functionally) independent FIs in involution. In this work, we have considered constrained autonomous conservative dynamical systems where the
constraint is a specified value of the total energy. This constraint defines in the configuration space a surface on which the trajectories of the dynamical system evolve. We proved a theorem which allows the systematic determination of autonomous and time-dependent exact QFIs of these constrained dynamical systems in terms of the symmetries of the kinetic metric; the latter being defined by the dynamical equations. Specifically, it is found that these QFIs are generated by CKVs and CKTs of the kinetic metric. Furthermore, there are three types of QFIs and specific formulae were given for their determination. It is to be noted that the time-dependent FIs appear for the first time in the literature of constrained systems.

In order to test the consequences and the validity
of the theorem, we have considered various applications from previous works in the topic. As a first application, we required the potential to admit a QFI of the second type, and we found a superintegrable potential of the Ermakov type and an integrable potential which for zero energy and zero QFI gives orbits which are circles. Next, in order to discuss the case of geodesics, we set the potential function equal to zero and determined the corresponding QFIs for the cases of null and timelike/spacelike constrained geodesics. The well-known QFIs of the unconstrained geodesics (see e.g. \cite{Katzin 1981, Tsamp 2020A, Tsamp 2020B}) are recovered as special cases. As an application in this area, we considered a general example which provides the various cases discussed in \cite{Dimakis 2019} using the Dirac-Bergmann formalism. We recovered all results of \cite{Dimakis 2019} and showed that one can integrate the constrained systems using exact FIs instead of the more complicated non-local FIs.

Theorem \ref{theorem.energy.constrained.systems} opens new directions in the study of the integrability of constrained dynamical systems. It would be interesting to generalize Theorem \ref{theorem.energy.constrained.systems} for other types of constraints, less trivial than the fixed energy constraint; and to reexamine well-known problems in Celestial Mechanics using  the constraint of specified energy with the possibility to obtain new QFIs.

\section*{Appendix}

\label{sec.solution}

Since the considered constrained dynamical system is autonomous, we should use the polynomial method described in \cite{Mits time} in order to solve the system of PDEs. According to this method, one assumes a general polynomial expression in the variable $t$ for both the quantities $K_{ab}(t,q)$ and $K_{a}(t,q)$, and replaces these expressions in the system of PDEs.

In particular, we assume that the CKT
\begin{equation}
K_{ab}(t,q)=C_{(0)ab}(q) + \sum_{P=1}^{m_{1}}C_{(P)ab}(q)\frac{t^{P}}{P} \label{eq.cons8.1}
\end{equation}%
where $C_{(P)ab}(q)$, $P=0,1,...,m_{1}$, is a sequence of second rank symmetric tensors, and the vector
\begin{equation}
K_{a}(t,q)=\sum_{M=0}^{m_{2}}L_{(M)a}(q)t^{M}  \label{eq.cons8.2}
\end{equation}%
where $L_{(M)a}(q)$, $M=0,1,...,m_{2}$, are arbitrary vectors.

We note that both powers $m_{1}$ and $m_{2}$ in the above polynomial expressions may be infinite.

Substituting (\ref{eq.cons8.1}) and (\ref{eq.cons8.2}) in equations (\ref{eq.cons6.1}) and (\ref{eq.cons6.2}), we obtain:
\begin{eqnarray}
X_{a}&=& X_{(0)a} +\sum_{P=1}^{m_{1}} X_{(P)a} \frac{t^{P}}{P} \label{eq.cons9.1} \\
\psi&=&\frac{1}{n} \left( \sum_{M=0}^{m_{2}} L_{(M)}{}^{a}{}_{;a} t^{M} +\sum_{P=1}^{m_{1}} C_{(P)}{}^{a}{}_{a} t^{P-1} \right) \label{eq.cons9.2}
\end{eqnarray}
where
\begin{equation}
X_{(P)a}= \frac{1}{n+2} \left( C_{(P)}{}^{b}{}_{b;a} +2C_{(P)}{}^{b}{}_{a;b} \right), \enskip P=0,1,...,m_{1}. \label{eq.cons10}
\end{equation}
Then, the PDEs (\ref{eq.cons5.1}) - (\ref{eq.cons5.4}) and (\ref{eq.cons7.1}) - (\ref{eq.cons7.2}) become:
\begin{align}
0=& C_{(0)(ab;c)}-X_{(0)(a}\gamma_{bc)} + \sum_{P=1}^{m_{1}} \left[ C_{(P)(ab;c)} -X_{(P)(a} \gamma_{bc)} \right] \frac{t^{P}}{P} \label{eq.cons11.1} \\
0=& \sum_{P=1}^{m_{1}} \left[ C_{(P)ab} -\frac{1}{n} C_{(P)}{}^{c}{}_{c} \gamma_{ab} \right] t^{P-1} +\sum_{M=0}^{m_{2}} \left[ L_{(M)(a;b)} -\frac{1}{n} L_{(M)}{}^{c}{}_{;c} \gamma_{ab} \right]t^{M} \label{eq.cons11.2} \\
0= & K_{,a} -2C_{(0)ab}V^{,b} -2(V-E_{0}) X_{(0)a} +\sum_{M=1}^{m_{2}} ML_{(M)a}t^{M-1} -2\sum_{P=1}^{m_{1}} \left[ C_{(P)ab} V^{,b} +(V-E_{0}) X_{(P)a} \right] \frac{t^{P}}{P} \label{eq.cons11.3} \\
0= & K_{,t} -\frac{2}{n}(V-E_{0}) \sum_{P=1}^{m_{1}} C_{(P)}{}^{a}{}_{a} t^{P-1} -\sum_{M=0}^{m_{2}} \left[ L_{(M)a}V^{,a} +\frac{2}{n}(V-E_{0}) L_{(M)}{}^{a}{}_{;a} \right] t^{M} \label{eq.cons11.4} \\
0= & \sum_{M=2}^{m_{2}} M(M-1)L_{(M)a}t^{M-2} +2\sum_{P=1}^{m_{1}} \left[ \frac{1}{n}(V-E_{0}) C_{(P)}{}^{b}{}_{b,a} -(V-E_{0})X_{(P)a} -C_{(P)ab}V^{,b} \right] t^{P-1} + \notag \\
& +\sum_{M=0}^{m_{2}} \left[ \left( L_{(M)b}V^{,b} \right)_{,a} +\frac{2}{n}(V-E_{0}) L_{(M)}{}^{b}{}_{;ba} \right] t^{M} \label{eq.cons11.5} \\
0= & 2\left( C_{(0)[a\left\vert c\right\vert }V^{,c}\right) _{;b]} +2\left[ (V-E_{0})X_{(0)[a} \right]_{;b]} -\sum_{M=1}^{m_{2}} ML_{(M)\left[a;b\right] }t^{M-1} +\notag \\
& +2\sum_{P=1}^{m_{1}} \left\{ \left( C_{(P)[a\left\vert c\right\vert }V^{,c} \right)_{;b]} + \left[ (V-E_{0})X_{(P)[a} \right]_{;b]} \right\} \frac{t^{P}}{P}. \label{eq.cons11.6}
\end{align}

Integrating equation (\ref{eq.cons11.4}), we find the scalar
\begin{equation}
K= \frac{2}{n}(V-E_{0}) \sum_{P=1}^{m_{1}} C_{(P)}{}^{a}{}_{a} \frac{t^{P}}{P} +\sum_{M=0}^{m_{2}} \left[ L_{(M)a}V^{,a} +\frac{2}{n}(V-E_{0}) L_{(M)}{}^{a}{}_{;a} \right] \frac{t^{M+1}}{M+1} +G(q) \label{eq.cons11.7}
\end{equation}
where $G(q)$ is an arbitrary function.

Replacing (\ref{eq.cons11.7}) in equation (\ref{eq.cons11.3}), we get the condition
\begin{eqnarray}
0&=& G_{,a} -2C_{(0)ab}V^{,b} -2(V-E_{0}) X_{(0)a} +\sum_{M=1}^{m_{2}} ML_{(M)a}t^{M-1} + \notag \\
&& +\sum_{M=0}^{m_{2}} \left[ \left( L_{(M)b}V^{,b} \right)_{,a} +\frac{2}{n}(V-E_{0}) L_{(M)}{}^{b}{}_{;ba} +\frac{2}{n}V_{,a} L_{(M)}{}^{b}{}_{;b}\right] \frac{t^{M+1}}{M+1} +\notag \\
&& +2\sum_{P=1}^{m_{1}} \left[ \frac{1}{n}(V-E_{0})C_{(P)}{}^{b}{}_{b,a} +\frac{1}{n}V_{,a}C_{(P)}{}^{b}{}_{b} -C_{(P)ab} V^{,b} -(V-E_{0}) X_{(P)a} \right] \frac{t^{P}}{P}. \label{eq.cons11.8}
\end{eqnarray}

We note that the integrability conditions (\ref{eq.cons11.5}) and (\ref{eq.cons11.6}) of the scalar $K$ are trivially satisfied, because the assumptions (\ref{eq.cons8.1}) and (\ref{eq.cons8.2}) bring the PDE (\ref{eq.cons5.4}) into the integrable form (\ref{eq.cons11.4}) whose integration gives directly the scalar $K(t,q)$ (see equation (\ref{eq.cons11.7}) ) in terms of the arbitrary function $G(q)$. Hence, the integrability conditions of $K$ are substituted by the integrability conditions $G_{,[ab]}=0$ of $G$, which in the case of the Euclidean plane $E^{2}$ produce the well-known second order Bertrand-Darboux equation \cite{Mits 2dpots, Darboux}. However, it can been checked that both equations (\ref{eq.cons11.5}) and (\ref{eq.cons11.6}) are always satisfied identically from the solutions of the other equations of the system.

Moreover, equation (\ref{eq.cons11.1}) implies that the second rank symmetric tensors $C_{(P)ab}(q)$ with $P=0,1,...,m_{1}$ are second order CKTs of the kinetic metric $\gamma_{ab}$ with associated vectors the quantities $X_{(P)a}$ given by (\ref{eq.cons10}). We conclude that the system of equations we have to solve consists of equations (\ref{eq.cons11.2}) and (\ref{eq.cons11.8}).

To solve equations (\ref{eq.cons11.2}) and (\ref{eq.cons11.8}), we consider the following cases:
\bigskip

\underline{\textbf{I. Case $\mathbf{m_{1}=m_{2}\equiv m}$}} (both $m_{1}$ and $m_{2}$ are finite)

In this case, equations (\ref{eq.cons11.2}) and (\ref{eq.cons11.8}) become:
\begin{align}
0=& \sum_{k=0}^{m-1} \left[ C_{(k+1)ab} -\frac{1}{n} C_{(k+1)}{}^{c}{}_{c} \gamma_{ab} +L_{(k)(a;b)} -\frac{1}{n} L_{(k)}{}^{c}{}_{;c} \gamma_{ab} \right]t^{k} +\left[ L_{(m)(a;b)} -\frac{1}{n} L_{(m)}{}^{c}{}_{;c} \gamma_{ab} \right]t^{m} \label{eq.cons12.1} \\
0=& G_{,a} -2C_{(0)ab}V^{,b} -2(V-E_{0}) X_{(0)a} +L_{(1)a} +\sum_{k=1}^{m-1} \left[ \left(L_{(k-1)b}V^{,b}\right)_{,a} -2C_{(k)ab} V^{,b} +k(k+1)L_{(k+1)a} + \right. \notag \\
& \left. +\frac{2}{n}(V-E_{0}) \left( C_{(k)}{}^{b}{}_{b,a} +L_{(k-1)}{}^{b}{}_{;ba} -nX_{(k)a} \right) +\frac{2}{n}V_{,a} \left( C_{(k)}{}^{b}{}_{b} +L_{(k-1)}{}^{b}{}_{;b} \right) \right] \frac{t^{k}}{k} +\left[ \left(L_{(m-1)b}V^{,b} \right)_{,a} - \right. \notag \\
& \left. -2C_{(m)ab} V^{,b} +\frac{2}{n}(V-E_{0}) \left( C_{(m)}{}^{b}{}_{b,a} +L_{(m-1)}{}^{b}{}_{;ba} -nX_{(m)a} \right) +\frac{2}{n}V_{,a} \left( C_{(m)}{}^{b}{}_{b} +L_{(m-1)}{}^{b}{}_{;b} \right) \right] \frac{t^{m}}{m} + \notag \\
& +\left[ \left(L_{(m)b}V^{,b} \right)_{,a} +\frac{2}{n}(V-E_{0}) L_{(m)}{}^{b}{}_{;ba} +\frac{2}{n}V_{,a} L_{(m)}{}^{b}{}_{;b} \right] \frac{t^{m+1}}{m+1}. \label{eq.cons12.2}
\end{align}

Equation (\ref{eq.cons12.1}) implies that $L_{(m)a}$ is a CKV of $\gamma_{ab}$ with conformal factor $\psi_{(m)}= \frac{1}{n} L_{(m)}{}^{a}{}_{;a}$ and the CKTs
\begin{equation}
C_{(k)ab} = f_{(k)} \gamma_{ab} -L_{(k-1)(a;b)}, \enskip k= 1, 2, ..., m \label{eq.cons13.1}
\end{equation}
where $f_{(k)}\equiv \frac{1}{n} \left[ C_{(k)}{}^{c}{}_{c} + L_{(k-1)}{}^{c}{}_{;c} \right]$.

From (\ref{eq.cons13.1}), we have:
\[
L_{(k-1)(a;b)}= f_{(k)} \gamma_{ab} -C_{(k)ab} \implies L_{(k-1)((a;b);c)}= f_{(k),(a} \gamma_{bc)} -\underbrace{C_{(k)(ab;c)}}_{=X_{(k)(a}\gamma_{bc)}} \implies L_{(k-1)((a;b);c)}= Y_{(k-1)(a}\gamma_{bc)}
\]
where the vector $Y_{(k-1)a}\equiv f_{(k),a} -X_{(k)a}$. Therefore, the vectors $L_{(k-1)a}$, $k=1,2,...,m$, produce the second order reducible CKTs $L_{(k-1)(a;b)}$ of $\gamma_{ab}$ with associated vectors
\begin{align*}
Y_{(k-1)a} =& f_{(k),a} -X_{(k)a} = f_{(k),a} -\frac{1}{n+2} \left( C_{(k)}{}^{b}{}_{b;a} +2C_{(k)}{}^{b}{}_{a;b} \right) \\
=& f_{(k),a} -\frac{1}{n+2} \left[ nf_{(k),a} -L_{(k-1)}{}^{b}{}_{;ba} +2\left( f_{(k)}\delta^{b}_{a} -\gamma^{bc} L_{(k-1)(c;a)} \right)_{;b} \right] \\
=& \frac{1}{n+2} \left[ L_{(k-1)}{}^{b}{}_{;ba} +2\gamma^{bc} L_{(k-1)(c;a);b} \right]
\end{align*}
which, as expected, are in accordance with the defining relation (\ref{eq.ConKT.1a}).

From equation (\ref{eq.cons12.2}), we find the following conditions\footnote{We have replaced the CKTs $C_{(k)ab}$ with $k=1,2,...,m$ from the relations (\ref{eq.cons13.1}).}:
\begin{eqnarray}
\left( L_{(m)b}V^{,b} \right)_{,a} &=& -2(V-E_{0})\psi_{(m),a} -2\psi_{(m)}V_{,a} \implies L_{(m)a}V^{,a} +2(V-E_{0})\psi_{(m)}= s= const \label{eq.cons13.3} \\
\left( L_{(m-1)b}V^{,b} \right)_{,a} &=& -2L_{(m-1)(a;b)}V^{,b} -2(V-E_{0}) Y_{(m-1)a} \label{eq.cons13.4} \\
\left( L_{(k-1)b}V^{,b} \right)_{,a} &=& -2L_{(k-1)(a;b)}V^{,b} -k(k+1)L_{(k+1)a} -2(V-E_{0})Y_{(k-1)a}, \enskip k=1,2, ..., m-1 \label{eq.cons13.5} \\
G_{,a}&=& 2C_{(0)ab}V^{,b} +2(V-E_{0}) X_{(0)a} -L_{(1)a}(m>0). \label{eq.cons13.6}
\end{eqnarray}
The notation $L_{(1)a}(m>0)$ indicates that the vector $L_{(1)a}$ exists only when $m>0$. If $m=0$, then $L_{(1)a}=0$.

From (\ref{eq.cons11.7}), we find that the scalar
\begin{align}
K=& \sum_{k=0}^{m-1} \left[ \frac{2}{n}(V-E_{0}) C_{(k+1)}{}^{a}{}_{a} +L_{(k)a}V^{,a} +\frac{2}{n}(V-E_{0}) L_{(k)}{}^{a}{}_{;a} \right] \frac{t^{k+1}}{k+1} + \notag \\ &+\left[ L_{(m)a}V^{,a} +\frac{2}{n}(V-E_{0}) L_{(m)}{}^{a}{}_{;a} \right] \frac{t^{m+1}}{m+1} +G(q). \label{eq.cons13.2}
\end{align}

Then, the QFI (\ref{eq.cons3}) is
\begin{align*}
I_{(m)} =& \left( \sum_{k=1}^{m} f_{(k)}\gamma_{ab} \frac{t^{k}}{k} -\sum_{k=1}^{m} L_{(k-1)(a;b)} \frac{t^{k}}{k} +C_{(0)ab} \right) \dot{q}^{a} \dot{q}^{b} +\sum_{k=0}^{m}t^{k}L_{(k)a}\dot{q}^{a} +\\
& +\left[ L_{(m)a}V^{,a} +2(V-E_{0})\psi_{(m)} \right] \frac{t^{m+1}}{m+1} +\sum_{k=1}^{m} \left[L_{(k-1)a}V^{,a} +2(V-E_{0})f_{(k)} \right] \frac{t^{k}}{k} +G(q) \\
=& \left( -\sum_{k=1}^{m} L_{(k-1)(a;b)} \frac{t^{k}}{k} +C_{(0)ab} \right) \dot{q}^{a} \dot{q}^{b} +\sum_{k=0}^{m}t^{k}L_{(k)a}\dot{q}^{a} +\left[ L_{(m)a}V^{,a} +2(V-E_{0})\psi_{(m)} \right] \frac{t^{m+1}}{m+1} +\\
& +\sum_{k=1}^{m} L_{(k-1)a}V^{,a} \frac{t^{k}}{k} +G(q) +\underbrace{\sum_{k=1}^{m}f_{(k)} \left[ \gamma_{ab} \dot{q}^{a} \dot{q}^{b} +2(V-E_{0}) \right] \frac{t^{k}}{k}}_{=0} \implies
\end{align*}
\begin{equation}
I_{(m)}= \left( -\sum_{k=1}^{m} L_{(k-1)(a;b)} \frac{t^{k}}{k} +C_{(0)ab} \right) \dot{q}^{a} \dot{q}^{b} +\sum_{k=0}^{m}t^{k}L_{(k)a}\dot{q}^{a} +s\frac{t^{m+1}}{m+1} +\sum_{k=1}^{m} L_{(k-1)a}V^{,a} \frac{t^{k}}{k} +G(q).
\label{eq.qfim}
\end{equation}
In (\ref{eq.qfim}), the index $(m)$ indicates the degree of the time-dependence of the coefficients, the quantities $C_{(0)ab}$ and $L_{(k)(a;b)}$ for $k=0,1,...,m-1$ are second order CKTs with associated vectors $X_{(0)a}$ and $Y_{(k)a}$, respectively, $L_{(m)a}$ is a CKV with conformal factor $\psi_{(m)}$, while the constant $s$, the vectors $L_{(k)a}$ and the function $G(q)$ satisfy the conditions (\ref{eq.cons13.3}) - (\ref{eq.cons13.6}).

In what follows, we show that the QFI $I_{(m)}$ given in (\ref{eq.qfim}) consists of two independent QFIs.

For small values of $m$, we have:

- For $m=0$.

The QFI is
\begin{equation}
I_{(0)}= C_{(0)ab}\dot{q}^{a}\dot{q}^{b} +L_{(0)a}\dot{q}^{a} +st +G(q) \label{eq.cons14}
\end{equation}
where $C_{(0)ab}$ is a CKT with associated vector $X_{(0)a}$, $L_{(0)a}$ is a CKV with conformal factor $\psi_{(0)}$ such that $L_{(0)a}V^{,a} +2(V-E_{0})\psi_{(0)}=s$ and the function $G(q)$ satisfies the condition $G_{,a}= 2C_{(0)ab}V^{,b} +2(V-E_{0}) X_{(0)a}$.

The QFI (\ref{eq.cons14}) consists of the independent FIs:
\begin{eqnarray*}
I_{(0,1)}&=& C_{(0)ab}\dot{q}^{a}\dot{q}^{b} +G(q) \\
I_{(0,2)}&=& L_{(0)a}\dot{q}^{a} +st.
\end{eqnarray*}

- For $m=1$.

The QFI is
\begin{equation}
I_{(1)} = \left( -tL_{(0)(a;b)} +C_{(0)ab} \right) \dot{q}^{a} \dot{q}^{b} +tL_{(1)a}\dot{q}^{a} +L_{(0)a}\dot{q}^{a} + s\frac{t^{2}}{2} +tL_{(0)a}V^{,a} +G(q) \label{eq.cons15}
\end{equation}
where $C_{(0)ab}$ and $L_{(0)(a;b)}$ are CKTs with associated vectors $X_{(0)a}$ and $Y_{(0)a}$, respectively, $L_{(1)a}$ is a CKV with conformal factor $\psi_{(1)}$ such that $L_{(1)a}V^{,a} +2(V-E_{0}) \psi_{(1)}= s$, while the vector $L_{(0)a}$ and the function $G(q)$ satisfy the conditions:
\begin{eqnarray*}
\left( L_{(0)b}V^{,b} \right)_{,a} &=& -2L_{(0)(a;b)}V^{,b} -2(V-E_{0}) Y_{(0)a} \\
G_{,a}&=& 2C_{(0)ab}V^{,b} +2(V-E_{0}) X_{(0)a} -L_{(1)a}.
\end{eqnarray*}

The QFI (\ref{eq.cons15}) consists of the independent FIs:
\begin{eqnarray*}
I_{(1,1)}&=& C_{(0)ab}\dot{q}^{a} \dot{q}^{b} +tL_{(1)a}\dot{q}^{a} +s\frac{t^{2}}{2} +G(q) \\
I_{(1,2)}&=& -tL_{(0)(a;b)} \dot{q}^{a} \dot{q}^{b} +L_{(0)a}\dot{q}^{a} + tL_{(0)a}V^{,a}.
\end{eqnarray*}

- For $m=2$.

The QFI is
\begin{eqnarray}
I_{(2)} &=& \left( -\frac{t^{2}}{2}L_{(1)(a;b)} -tL_{(0)(a;b)} + C_{(0)ab}\right) \dot{q}^{a} \dot{q}^{b} +t^{2}L_{(2)a}\dot{q}^{a} + t L_{(1)a}\dot{q}^{a} + L_{(0)a}\dot{q}^{a} + \notag \\
&& + s\frac{t^{3}}{3} + \frac{t^{2}}{2}L_{(1)a}V^{,a} +t L_{(0)a}V^{,a} + G(q) \label{eq.cons16}
\end{eqnarray}
where $C_{(0)ab}$ and $L_{(k)(a;b)}$ for $k=0,1$ are CKTs with associated vectors $X_{(0)a}$ and $Y_{(k)a}$, respectively, $L_{(2)a}$ is a CKV with conformal factor $\psi_{(2)}$ such that $L_{(2)a}V^{,a} +2(V-E_{0})\psi_{(2)} =s$, while the vectors $L_{(k)a}$ and the function $G(q)$ satisfy the conditions:
\begin{eqnarray*}
\left( L_{(1)b}V^{,b} \right)_{,a} &=& -2L_{(1)(a;b)}V^{,b} -2(V-E_{0}) Y_{(1)a} \\
\left( L_{(0)b}V^{,b} \right)_{,a} &=& -2L_{(0)(a;b)}V^{,b} -2L_{(2)a} -2(V-E_{0})Y_{(0)a} \\
G_{,a}&=& 2C_{(0)ab}V^{,b} +2(V-E_{0}) X_{(0)a} -L_{(1)a}.
\end{eqnarray*}

The QFI (\ref{eq.cons16}) consists of the independent FIs:
\begin{eqnarray*}
I_{(2,1)} &=& \left( -\frac{t^{2}}{2}L_{(1)(a;b)} + C_{(0)ab} \right) \dot{q}^{a} \dot{q}^{b} + t L_{(1)a}\dot{q}^{a} + \frac{t^{2}}{2} L_{(1)a}V^{,a} +G(q) \\
I_{(2,2)} &=& -tL_{(0)(a;b)} \dot{q}^{a} \dot{q}^{b} + t^{2}L_{(2)a}\dot{q}^{a}+ L_{(0)a}\dot{q}^{a} + s\frac{t^{3}}{3} +t L_{(0)a}V^{,a}.
\end{eqnarray*}

- For $m=3$.

\begin{eqnarray}
I_{(3)} &=& \left( - \frac{t^{3}}{3} L_{(2)(a;b)} - \frac{t^{2}}{2}L_{(1)(a;b)} - t L_{(0)(a;b)} + C_{(0)ab} \right) \dot{q}^{a} \dot{q}^{b} +t^{3} L_{(3)a}\dot{q}^{a} + t^{2}L_{(2)a}\dot{q}^{a} + t L_{(1)a}\dot{q}^{a}+ \notag \\
&& + L_{(0)a}\dot{q}^{a} + s\frac{t^{4}}{4} + \frac{t^{3}}{3} L_{(2)a}V^{,a} + \frac{t^{2}}{2} L_{(1)a}V^{,a} +tL_{(0)a}V^{,a} + G(q) \label{eq.cons17}
\end{eqnarray}
where $C_{(0)ab}$ and $L_{(k)(a;b)}$ for $k=0,1,2$ are CKTs with associated vectors $X_{(0)a}$ and $Y_{(k)a}$, respectively, $L_{(3)a}$ is a CKV with conformal factor $\psi_{(3)}$ such that $L_{(3)a}V^{,a} +2(V-E_{0})\psi_{(3)}=s$, while the vectors $L_{(k)a}$ and the function $G(q)$ satisfy the conditions:
\begin{eqnarray*}
\left( L_{(2)b}V^{,b} \right)_{,a} &=& -2L_{(2)(a;b)}V^{,b} -2(V-E_{0}) Y_{(2)a} \\
\left( L_{(1)b}V^{,b} \right)_{,a} &=& -2L_{(1)(a;b)}V^{,b} -6L_{(3)a} -2(V-E_{0})Y_{(1)a} \\
\left( L_{(0)b}V^{,b} \right)_{,a} &=& -2L_{(0)(a;b)}V^{,b} -2L_{(2)a} -2(V-E_{0})Y_{(0)a} \\
G_{,a}&=& 2C_{(0)ab}V^{,b} +2(V-E_{0}) X_{(0)a} -L_{(1)a}.
\end{eqnarray*}

The QFI (\ref{eq.cons17}) consists of the independent FIs:
\begin{eqnarray*}
I_{(3,1)} &=& \left( - \frac{t^{2}}{2} L_{(1)(a;b)} + C_{(0)ab} \right) \dot{q}^{a} \dot{q}^{b} + t^{3} L_{(3)a}\dot{q}^{a} + t L_{(1)a}\dot{q}^{a} + s\frac{t^{4}}{4} + \frac{t^{2}}{2} L_{(1)a}V^{,a} + G(q) \\
I_{(3,2)} &=& \left( - \frac{t^{3}}{3} L_{(2)(a;b)} - tL_{(0)(a;b)} \right)\dot{q}^{a} \dot{q}^{b}+ t^{2}L_{(2)a}\dot{q}^{a} + L_{(0)a}\dot{q}^{a} +
\frac{t^{3}}{3} L_{(2)a}V^{,a} +t L_{(0)a}V^{,a}.
\end{eqnarray*}

By mathematical induction, it is proved that the QFI $I_{(m)}$ consists of the following two independent QFIs:

a.
\[
I_{(\ell)1}= \left( - \sum_{k=1}^{\ell} \frac{t^{2k}}{2k} L_{(2k-1)(a;b)} + C_{(0)ab}\right) \dot{q}^{a}\dot{q}^{b} + \sum_{k=1}^{\ell} t^{2k-1} L_{(2k-1)a}\dot{q}^{a} + \sum_{k=1}^{\ell} \frac{t^{2k}}{2k} L_{(2k-1)a}V^{,a}+ G(q)
\]
where $C_{(0)ab}$ and $L_{(2k-1)(a;b)}$ for $k=1,2,...,\ell$ are CKTs with associated vectors $X_{(0)a}$ and $Y_{(2k-1)a}$, respectively, and the vectors $L_{(2k-1)a}$ and the function $G(q)$ satisfy the conditions:
\begin{eqnarray*}
\left( L_{(2\ell-1)b}V^{,b} \right)_{,a} &=& -2L_{(2\ell-1)(a;b)}V^{,b} -2(V-E_{0}) Y_{(2\ell-1)a} \\
\left( L_{(2k-1)b}V^{,b} \right)_{,a} &=& -2L_{(2k-1)(a;b)}V^{,b} -2k(2k+1)L_{(2k+1)a} -2(V-E_{0})Y_{(2k-1)a}, \enskip k=1,2, ..., \ell-1 \\
G_{,a}&=& 2C_{(0)ab}V^{,b} +2(V-E_{0}) X_{(0)a} -L_{(1)a}(\ell>0).
\end{eqnarray*}

b.
\[
I_{(\ell)2}= \sum_{k=0}^{\ell} \left( -\frac{t^{2k+1}}{2k+1} L_{(2k)(a;b)}\dot{q}^{a} \dot{q}^{b} +t^{2k} L_{(2k)a}\dot{q}^{a}+ \frac{t^{2k+1}}{2k+1} L_{(2k)a}V^{,a} \right)
\]
where $L_{(2k)(a;b)}$ for $k=0,1,...,\ell$ are CKTs with associated vectors $Y_{(2k)a}$ and the vectors $L_{(2k)a}$ satisfy the conditions:
\begin{eqnarray*}
\left( L_{(2\ell)b}V^{,b} \right)_{,a} &=& -2L_{(2\ell)(a;b)}V^{,b} -2(V-E_{0}) Y_{(2\ell)a} \\
\left( L_{(2k)b}V^{,b} \right)_{,a} &=& -2L_{(2k)(a;b)}V^{,b} -2(k+1)(2k+1)L_{(2k+2)a} -2(V-E_{0})Y_{(2k)a}, \enskip k=0,1,...,\ell-1.
\end{eqnarray*}

\underline{\textbf{II. Case $\mathbf{m_{1} \neq m_{2}}$}.} ($m_{1}$ or $m_{2}$ may be infinite)

We find QFIs that are subcases of those found in \textbf{Case I} and \textbf{Case III} below.

\underline{\textbf{III. Both $\mathbf{m_{1}}$ and $\mathbf{m_{2}}$ are infinite.}}

In this case, the only non-trivial solution is for $K_{ab}= e^{\lambda t} C_{ab}(q)$ and $K_{a}= e^{\mu t} L_{a}(q)$, where $C_{ab}$ is a second order CKT with associated vector $B_{a}(q)$ and the constants $\lambda \mu \neq0$.

Then, the scalar (\ref{eq.cons11.7}) is
\begin{equation}
K= \frac{e^{\mu t}}{\mu} \left[ L_{a}V^{,a} +\frac{2}{n}(V-E_{0})L^{a}{}_{;a} \right] +\frac{2}{n}(V-E_{0}) C^{a}{}_{a} e^{\lambda t} +G(q) \label{eq.cons18}
\end{equation}
and the PDEs (\ref{eq.cons11.2}), (\ref{eq.cons11.8}) become:
\begin{eqnarray}
0 &=& \lambda e^{\lambda t} \left( C_{ab} -\frac{1}{n} C^{c}{}_{c}\gamma_{ab} \right) +e^{\mu t} \left( L_{(a;b)} -\frac{1}{n}L^{c}{}_{;c}\gamma_{ab} \right) \label{eq.cons18.1} \\
0 &=& G_{,a} +\frac{e^{\mu t}}{\mu} \left[ \left(L_{b}V^{,b} \right)_{,a} +\frac{2}{n}(V-E_{0})L^{b}{}_{;ba} +\frac{2}{n}V_{,a}L^{b}{}_{;b} +\mu^{2} L_{a}\right] +\notag \\ && +2e^{\lambda t} \left[ \frac{1}{n}(V-E_{0}) C^{b}{}_{b,a} +\frac{1}{n}V_{,a} C^{b}{}_{b} -C_{ab}V^{,b} -(V-E_{0}) B_{a}\right].
\label{eq.cons18.2}
\end{eqnarray}

We consider the following subcases:

1) For $\lambda \neq \mu$.

Equation (\ref{eq.cons18.1}) implies that $C_{ab}= f\gamma_{ab}$ is a second order gradient CKT with associated vector $B_{a}=f_{,a}$, where $f(q)$ is a smooth function, and $L_{a}$ is a CKV with conformal factor $\rho(q)$.

From the condition (\ref{eq.cons18.2}), we get\footnote{Recall that by contracting with $\gamma^{ab}$ the relations $C_{ab}=f\gamma_{ab}$ and $L_{(a;b)}= \rho\gamma_{ab}$, we find $f=\frac{1}{n}C^{a}{}_{a}$ and $\rho= \frac{1}{n}L^{a}{}_{;a}$, respectively.}:
\[
G=const\equiv 0, \enskip L_{a}= -\frac{1}{\mu^{2}} \left[ L_{b}V^{,b} +2(V-E_{0})\rho \right]_{,a}.
\]

The scalar (\ref{eq.cons18}) is written
\[
K=  \frac{e^{\mu t}}{\mu} \left[ L_{a}V^{,a} +2(V-E_{0})\rho \right] +2(V-E_{0})f e^{\lambda t} +G(q).
\]

The QFI reduces to the LFI
\begin{eqnarray*}
I_{e}(\lambda \neq \mu)&=& \underbrace{e^{\lambda t}f \gamma_{ab}\dot{q}^{a}\dot{q}^{b} +2(V-E_{0})f e^{\lambda t}}_{=0} +e^{\mu t}L_{a}\dot{q}^{a} +\frac{e^{\mu t}}{\mu} \left[ L_{a}V^{,a} +2(V-E_{0})\rho \right] \\
&=& e^{\mu t}L_{a}\dot{q}^{a} +\frac{e^{\mu t}}{\mu} \left[ L_{a}V^{,a} +2(V-E_{0})\rho \right]
\end{eqnarray*}
where $L_{a}= -\frac{1}{\mu^{2}} \left[ L_{b}V^{,b} +2(V-E_{0})\rho \right]_{,a}$ is a gradient CKV with conformal factor $\rho(q)$.

2) For $\lambda = \mu$.

In this case, equation (\ref{eq.cons18.1}) implies that the CKT $\lambda C_{ab}= g\gamma_{ab} -L_{(a;b)}$, where $g(q)$ is a smooth function. Then, $L_{(a;b)}$ is a reducible CKT with associated vector $Y_{a}\equiv g_{,a} -\lambda B_{a}$.

From the remaining condition (\ref{eq.cons18.2}), we get\footnote{Recall that by contracting the relation $\lambda C_{ab}= g\gamma_{ab} -L_{(a;b)}$, we find $g= \frac{1}{n}\left( \lambda C^{a}{}_{a} +L^{a}{}_{;a} \right)$.}:
\[
G=const\equiv 0, \enskip \left( L_{b}V^{,b} \right)_{,a}= -2L_{(a;b)}V^{,b} -\lambda^{2}L_{a} -2(V-E_{0})Y_{a}.
\]

The scalar (\ref{eq.cons18}) is written
\[
K= \frac{e^{\lambda t}}{\lambda} \left[ L_{a}V^{,a} +2(V-E_{0})g \right] +G(q).
\]

The QFI is\footnote{We multiply with the non-zero constant $\lambda$.}
\begin{eqnarray*}
I_{e}(\lambda=\mu) &=& e^{\lambda t} \left( -L_{(a;b)} \dot{q}^{a} \dot{q}^{b} +\lambda L_{a}\dot{q}^{a} +L_{a}V^{,a} \right) +g e^{\lambda t} \underbrace{\left[ \gamma_{ab}\dot{q}^{a} \dot{q}^{b} +2(V-E_{0}) \right]}_{=0} \\
&=& e^{\lambda t} \left( -L_{(a;b)} \dot{q}^{a} \dot{q}^{b} +\lambda L_{a}\dot{q}^{a} +L_{a}V^{,a} \right)
\end{eqnarray*}
where $L_{(a;b)}$ is a reducible CKT with associated vector $Y_{a}$ such that $\left( L_{b}V^{,b} \right)_{,a}= -2L_{(a;b)}V^{,b} -\lambda^{2}L_{a} -2(V-E_{0})Y_{a}$.

We note that the LFI $I_{e}(\lambda \neq \mu)$ is derived from $I_{e}(\lambda=\mu)$ in the case that $L_{a}$ is a CKV.
\bigskip

The above complete the proof of Theorem \ref{theorem.energy.constrained.systems}.

\end{document}